\documentclass[a4paper,11pt]{article}
\pdfoutput=1 % if your are submitting a pdflatex (i.e. if you have
\usepackage [latin1]{inputenc}
\usepackage{jinstpub} % for details on the use of the package, please
\usepackage{rotating}
\usepackage{xcolor}
\usepackage{siunitx}

\usepackage{xspace}
\newcommand{\pmt}{PM\xspace}
\newcommand{\pmts}{PMs\xspace}
\usepackage{newtxtext}
\usepackage{newtxmath}

\title{\boldmath Light Detection System and Time Resolution of the NA62~RICH}

\author[a,b]{G.~Anzivino,}
\author[a,b]{M.~Barbanera,}
\author[c,d]{A.~Bizzeti,}
\author[a,b]{F.~Brizioli,}
\author[d]{F.~Bucci,}
\author[e,d]{A.~Cassese,}
\author[b]{P.~Cenci,}
\author[d]{R.~Ciaranfi,}
\author[b,2]{V.~Duk,}
\author[f]{J.~Engelfried,}
\author[f,g]{N.~Estrada-Tristan,}
\author[e,d]{E.~Iacopini,}
\author[a,b]{E.~Imbergamo,}
\author[e,d]{G.~Latino,}
\author[e,d,1]{M.~Lenti\note{Corresponding author.}\note{Funded by the EU Horizon 2020 research and innovation programme (Marie 
Sk\l{}odowska-Curie
grant No 701386).}
,}
\author[a,b]{R.~Lollini,}
\author[b]{P.~Lubrano,}
\author[a,b,h]{R.~Piandani,}
\author[b]{M.~Pepe,}
\author[b]{M.~Piccini,}
\author[a,b,i]{A.~Sergi,}
\author[e,d]{M.~Turisini,}
\author[j,k]{R.~Volpe}
\affiliation[a]{Dipartimento di Fisica e Geologia dell'Università di Perugia,\\I-06123 Perugia, Italy}
\affiliation[b]{Sezione dell'INFN di Perugia,\\I-06123 Perugia, Italy}
\affiliation[c]{Dipartimento di Scienze Fisiche, Informatiche e Matematiche dell'Università di Modena e Reggio Emilia,\\I-41125 Modena, Italy}
\affiliation[d]{Sezione dell'INFN di Firenze,\\I-50019 Sesto Fiorentino, Italy}
\affiliation[e]{Dipartimento di Fisica e Astronomia dell'Università  di Firenze,\\ I-50019 Sesto Fiorentino, Italy}
\affiliation[f]{Instituto de Física, Universidad Autónoma de San Luis Potosí, San Luis Potosí, Mexico}
\affiliation[g]{Also at Universidad de Guanajuato, Guanajuato, Mexico}
\affiliation[h]{Now at: Institut fur Experimentelle Teilchenphysik, Karlsruhe Institute of Technology (KIT),76131 Karlsruhe, Germany}
\affiliation[i]{Now at: University of Birmingham, Birmingham, United Kingdom}
\affiliation[j]{Université Catholique de Louvain, Louvain-La-Neuve, Belgium}
\affiliation[k]{Now at: Faculty of Mathematics, Physics and Informatics, Comenius University, 842 48, Bratislava, Slovakia}

\emailAdd{massimo.lenti@fi.infn.it}

\abstract{
A large RICH detector is used 
in NA62
to suppress the muon contamination in the charged pion %selection 
sample
by a factor of 100 in the momentum range between 15 and 35 GeV/c.
Cherenkov light is 
collected 
by 1952 photomultipliers %(PMs) 
placed at the upstream end. 
In this 
paper 
the characterization of the %PMs 
photomultipliers 
and the dedicated Frontend %(FE) 
and Data Acquisition %(DAQ) 
electronics are described,
the time resolution 
and the light detection efficiency measurement 
are %is 
presented.}


\begin{document}
\maketitle
\flushbottom

%%%%%%%%%%%%%%%%%%
\section{Introduction}
\label{sec:intro}
%%%%%%%%%%%%%%%%%%
\begin{figure}[t]
  \begin{center}
 \includegraphics[width=1.08\textwidth]{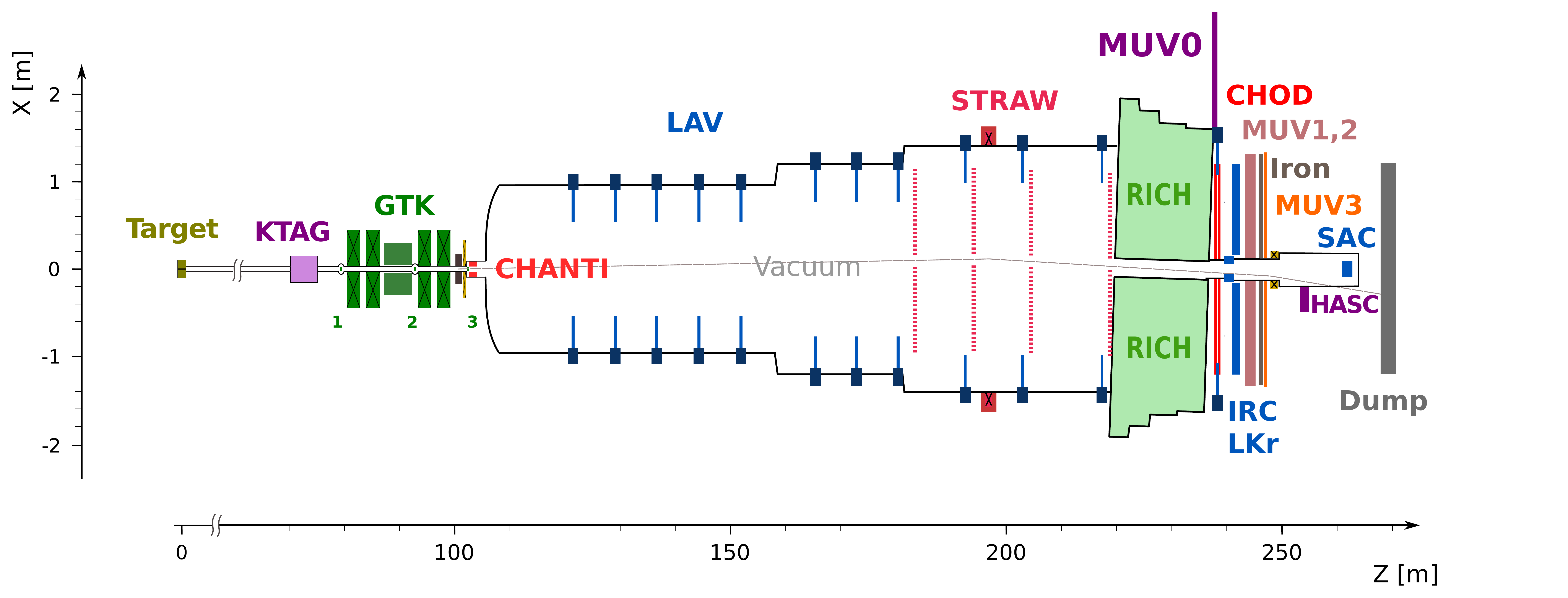}
    \caption{Schematic top view of the NA62 beam line and detector. Dipole magnets are displayed as boxes with superimposed crosses. Also shown is the trajectory of a beam particle 
    in vacuum which crosses all the detector apertures, thus avoiding interactions with material. A dipole magnet between MUV3 and SAC deflects the beam particles out of the SAC acceptance.}
    \label{fig:layoutxz17}
  \end{center}
\end{figure}
NA62~\cite{proposal} is a fixed target experiment located in the North Area of CERN and dedicated to study charged kaon decays, in 
particular the very rare decay $K^+\rightarrow\pi^+\nu\bar\nu$. 
%BRIGITTE'S TEXT.
The NA62 beam line and detector are sketched in figure~\ref{fig:layoutxz17},
while a detailed description 
% of them 
can be found in~\cite{na62det}.  
%The beam line defines
%the Z-axis of the experiment's right-handed laboratory coordinate system.  The
%origin is the kaon production target, and beam particles travel in the
%positive Z-direction.  The Y-axis is vertical (positive up), and the X-axis is
%horizontal (positive left).

%The kaon production target is a 40-cm long beryllium rod.  
A 400-GeV proton beam extracted from the CERN Super Proton Synchrotron (SPS) 
impinges on 
%the target 
a beryllium target
in spills of three seconds effective duration.    Typical
intensities during data-taking range from  $1.7\times10^{12}$ to $1.9\times10^{12}$
protons per pulse. 
% ($ppp$).  
The resulting secondary hadron beam of positively
charged particles is comprised of 70\% $\pi^+$, 23\% protons, 
6\% $K^+$,
and 1$\%$ $\mu^{+}$,
with a nominal momentum of 75 GeV/c (1\% rms momentum bite).

Beam particles are characterized by a differential Cherenkov counter (KTAG)
and a three-station silicon pixel array (Gigatracker, GTK).  The KTAG uses N$_2$ gas 
at 1.75~bar pressure (contained in a 5-m long vessel) and is read out by photomultiplier tubes grouped in eight
sectors.  It tags incoming kaons with 70-ps time resolution.  The GTK stations
are located before, between, and after two pairs of dipole magnets (a beam
achromat), forming a  spectrometer that measures beam particle momentum, direction,
and time with resolutions of 0.15 GeV/$c$, 16 $\mu$rad, and 100 ps,
respectively.  The typical beam particle rate at the third GTK station (GTK3)
is about 450~MHz.

%This last station is immediately preceded by a 1-m thick, variable
%aperture steel collimator.  Its inner aperture is typically set at
%66~mm $\times$ 33~mm, and its outer dimensions are about 15~cm.  It serves as
%a partial shield against hadrons produced by upstream $K^+$ decays.
%
GTK3 marks the beginning of a 117~m-long vacuum tank. The first 80~m of the
tank define a fiducial volume (FV) in which 13\% of the kaons decay.  
%The beam
%has a  rectangular transverse profile of 52 $\times$ 24 mm$^2$ and a divergence
%of 0.11~mrad (rms) in each plane at the FV entrance.

The time, momentum, and direction of charged 
%daughters of kaon decays-in-flight
decay products
are measured by a magnetic spectrometer (STRAW), a ring-imaging Cherenkov
counter (RICH), and two scintillator hodoscopes (CHOD and NA48-CHOD).
The STRAW, comprised of two pairs of straw chambers on either side of a
dipole magnet,  measures momentum-vectors with a resolution of
$\sigma_p / p$ between 0.3\% and 0.4\%.  
The RICH
%, filled with neon at atmospheric pressure, 
must tag the decay particles with a timing precision of
better than 100~ps and provides particle identification.  The CHOD, a matrix of
tiles read out by SiPMs, and the NA48-CHOD, composed of two orthogonal planes
of scintillating slabs, 
%reused from the NA48 experiment, 
are used for
triggering and timing, providing a time measurement with 200-ps resolution.
 
Other sub-detectors suppress decays into photons or into multiple charged
particles (electrons, pions or muons) or provide complementary particle
identification.  Six stations of plastic scintillator bars (CHANTI) detect,
with 99\% efficiency and 1~ns time resolution, extra activity, including
inelastic interactions in  GTK3.  Twelve stations of ring-shaped
electromagnetic calorimeters (LAV1 to LAV12), made of lead-glass blocks,
surround the vacuum tank and downstream sub-detectors to achieve hermetic
acceptance for photons emitted by $K^+$ decays in the FV at polar angles
between 10 and 50~mrad.  A 27 radiation-length thick, quasi-homogeneous liquid
krypton electromagnetic calorimeter (LKr) detects photons from $K^+$  decays
emitted at angles between 1 and 10~mrad.  The LKr also complements the RICH
for particle identification.  
%Its energy resolution in NA62 conditions is
%$\sigma_E / E = 1.4\%$ for energy deposits of 25~GeV.  Its spatial and time
%resolutions are 1~mm and between 0.5 and 1~ns, respectively, depending on the
%amount and type of energy released.  
Two hadronic iron/scintillator-strip
sampling calorimeters (MUV1,2) and an array of scintillator tiles located
behind 80 cm of iron (MUV3)  supplement the pion/muon identification system. 
%MUV3 has a time resolution of 400~ps.  
A lead/scintillator shashlik calorimeter
(IRC) located in front of the LKr
%, covering an annular region between 65 and 135~mm from the Z axis, 
and a similar detector (SAC) placed on the Z axis at
the downstream end of the apparatus, ensure the detection of photons down to
zero degrees in the forward direction.  Additional counters (MUV0, HASC)
installed at optimized locations provide nearly hermetic coverage for charged
particles produced in multi-track kaon decays.

%All detectors are read out with TDCs, except for LKr and MUV1, 2, which are
%read out with 14-bit FADCs.  The IRC and SAC are read out with both.  All
%TDCs are mounted on custom-made (TEL62) boards, except for GTK and STRAW,
%which each have specialized TDC boards.  TEL62 boards both read out data and
%provide trigger information.  A dedicated processor interprets calorimeter
%signals for triggering~\cite{l0calo} . 
%A dedicated board combines logical signals (primitives) from the RICH, CHOD,
%NA48-CHOD, LKr, LAV, and MUV3 into a low-level trigger (L0) whose decision is
%dispatched to sub-detectors for data readout~\cite{l0tp}. 
%A software trigger (L1) exploits reconstruction
%algorithms similar to those used offline with data from KTAG, LAV, and
%STRAW to further cull the data before storing it on disk~\cite{na62det}.
%
%The analysis also uses data taken with a minimum-bias L0 trigger (control trigger)
%based on NA48-CHOD information downscaled by a factor of 400 to measure efficiencies and estimate backgrounds. 

%END of BRIGITTE'S TEXT
%~\cite{fulldetector}. 
%% and a view of the NA62 experiment layout is shown in figure~\ref{fig:na62layout}.
The RICH detector has a fundamental role in the search for  the $K^+\rightarrow\pi^+\nu\bar\nu$ decay~\cite{roberta}.
It must distinguish pions from muons, 
provide a precise pion time in a high particle rate environment (10 MHz)
to match the kaon in the very intense beam (750 MHz); 
furthermore the RICH time is used as reference for the Level-0 trigger of the experiment. 
A first study of $K^+\rightarrow\pi^+\nu\bar\nu$ decay, based on data collected in 2016, was published in~\cite{na62result2016},
while the analysis of data collected in 2017 was presented in~\cite{na62result2017}, where the details of background suppression 
are discussed and the importance of the RICH detector is highlighted.

%%%%%%%%%%%%%%%%%%
\section{RICH detector}
\label{sec:rich}
%%%%%%%%%%%%%%%%%%
The process $K^+\rightarrow\mu^+\nu$ is the $K^+$ decay with the largest branching ratio ($\sim63.6\%$),
about 10 orders of magnitude larger than 
that of
the signal $K^+\rightarrow\pi^+\nu\bar\nu$. 
The suppression of this 
background is achieved by kinematics %methods 
and by using the very different stopping power of muons and 
charged pions in
% the 
calorimeters. A further suppression factor of 100 is obtained by means of 
% a Cherenkov detector, 
the RICH.  
Neon gas at atmospheric pressure has been chosen as radiator having the appropriate refractive index and because it guarantees good light transmission and low chromatic dispersion. The 12.5 GeV/c Cherenkov threshold for pions is well suited for the 15 GeV/c lower bound of the NA62 operating momentum range. On the other hand, given the low emission rate per length of Cherenkov photons in neon, a long radiator is needed. 
\begin{figure}[htbp]
\centering % \begin{center}/\end{center} takes some additional vertical space
\includegraphics[width=1.\textwidth,trim=0 0 0 0,clip]{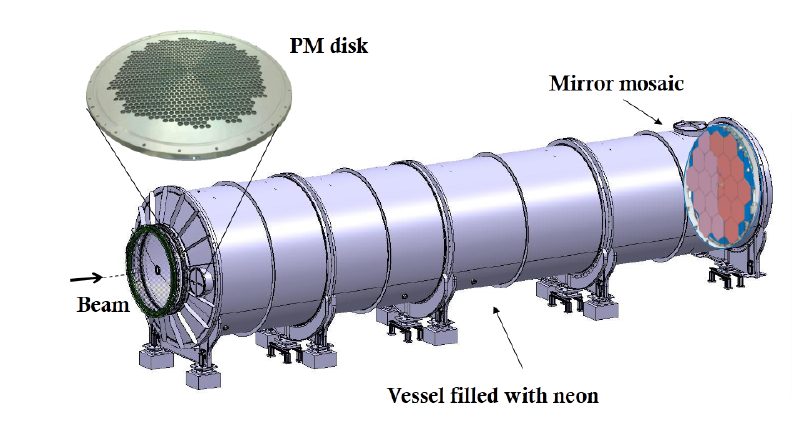}
% "\includegraphics" from the "graphicx" permits to crop (trim+clip)
% and rotate (angle) and image (and much more)
%% \put(-271,165){\colorbox{gray!0}{\makebox(1,10){}}}
%\put(-273,164){\colorbox{gray!0}{\textcolor{green}{s}}}
%% \put(-270,164){\textcolor{teal}{\textbf{s}}}
%% \put(-286,165){\colorbox{gray!0}{\makebox(35,10){}}}
%% \put(-286,164){\textcolor{teal}{\textbf{PM disk}}}
%% \put(-160,165){\colorbox{gray!0}{\makebox(65,10){}}}
%% \put(-160,166){\textcolor{teal}{\textbf{Mirror mosaic}}}
%% \put(-425,64){\colorbox{gray!0}{\makebox(25,5){}}}
%% \put(-425,64){\textcolor{teal}{\textbf{Beam}}}
%% \put(-210,6){\colorbox{gray!0}{\makebox(105,5){}}}
%% \put(-210,6){\textcolor{teal}{\textbf{Vessel filled with neon}}}
\caption{\label{fig:rich} The RICH detector. The hadron beam enters from the left and travels %in vacuum 
through a beam pipe which crosses the detector. A zoom on one of the two disks 
accommodating the %light sensors (photomultipliers, 
PMs is shown on the left; the mirror mosaic inside the neon %container ("
vessel %") 
is visible on the right.}
\end{figure}

A detailed description of the RICH detector can be found in~\cite{mirrors}. The RICH neon vessel is a 17 m long vacuum proof tank, made of steel and composed of 4 cylindrical sections of decreasing diameter. A beam pipe, 168 mm in diameter, crosses the RICH detector to allow undecayed particles to travel in vacuum down to the experiment beam dump. A steel conical cap connects the upstream window with the first cylindrical section and accommodates the photomultipliers (PMs) placed outside the particle acceptance in two disks containing 976 PMs each. Cherenkov light is reflected by a mosaic of 20 spherical mirrors with 17 m focal length, placed at the downstream end of the vessel and described in detail in~\cite{mirrors}; the mirror alignment during data taking is discussed in~\cite{alignment}. The total neon volume is about 200 m$^3$: fresh neon is injected inside the container after it has been fully evacuated, then sealed. No purification or recirculation system is present. Two prototype detectors were built and tested in hadron beams to demonstrate the performance of the proposed layout: the results of these tests have been published~\cite{protoa,protob}.  
A 
% schematic 
pictorial
view of the RICH detector is shown in figure~\ref{fig:rich}.

%%%%%%%%%%%%%%%%%%
%%%%%%%%%%%%%%%%%%
%\section{The photomultiplier}
\section{The light detection system}
%\section{Photomultipliers}
\label{sec:PMs}
%%%%%%%%%%%%%%%%%%
%%%%%%%%%%%%%%%%%%
The simulation of the shape and granularity of the photo-detection system showed that the optimal performance can be achieved with a pixel pitch not larger than 18 mm, given the dual mirror orientation, the radiator refractive index and the signal kinematics.

A compact size PM (Hamamatsu R7400 U-03) with cathode sensitivity ranging from visible to near UV and high multiplication gain was chosen as photo-detector as the result of a beam test~\cite{protoa}. 
% The PM has a circular active surface of 8 mm diameter, 
% an outer width of 16 mm and a height of 16 mm. 
The PM has a cylindrical shape with an outer diameter of 16 mm,
a height of 16 mm and a circular active surface of 8 mm diameter.
% The PM has a 
The transit time spread is  0.28 ns. 
The characteristics of the chosen PM are summarised in table~\ref{tab:pmcharcter}.

\begin{table}[htbp]
\centering
\caption{\label{tab:pmcharcter} Main characteristics of Hamamatsu R7400U-03 photomultiplier at HV=800 V.}
\smallskip
\begin{tabular}{|c|c|}
\hline
PM & R7400U-03 \\ \hline
type & head on; metal package tube for UV-visible range \\ \hline
diameter & 16 mm \\ \hline
active diameter & 8 mm \\ \hline
min $\lambda$ & 185 nm \\ \hline
max $\lambda$ & 650 nm \\ \hline
peak sensitivity & 420 nm \\ \hline
cathode radiant sensitivity & 62 mA/W \\ \hline
window & UV glass \\ \hline
cathode type & bialkali \\ \hline
cathode luminous sensitivity & 70 $\mu$A/lm \\ \hline
anode luminous sensitivity & 50 A/lm \\ \hline
gain & $7.0\times 10^5$ \\ \hline
dark current after 30 minutes & 0.2 nA \\ \hline
rise time & 0.78 ns \\ \hline
transit time & 5.4 ns \\ \hline
transit time spread & 0.28 ns \\ \hline
number of dynodes & 8 \\ \hline
%Applied Voltage & 800 V \\ \hline
\end{tabular}
\end{table}

%%%%%%%%%%%%%%%%%%
%\subsection{Characteristics} 
\subsection{The photomultipliers} 
%% \subsection{PM single photon response}
\label{subsec:PMcharacterization}
%%%%%%%%%%%%%%%%%%
The R7400U-03 spectral response was measured in laboratory for several units and for few photon wavelengths since a systematic characterization in the whole Cherenkov spectrum would have required dedicated instrumentation. 
Those measurements 
% confirmed 
showed
that the quantum efficiency (QE) is about    20\% at 400 nm for PMs with cathode luminous sensitivity at the level of 60 $\mu$A lm$^{-1}$.
Moreover, the efficiency was found to be proportional to the cathode luminous sensitivity quoted by
manufacturer, hence 
the 
1952 PMs with the best cathode luminous sensitivity were installed in the RICH detector and the remaining PMs ($\sim$15\%) 
% have been 
were
left as spares.

\begin{figure}[htbp]
\centering % \begin{center}/\end{center} takes some additional vertical space
\includegraphics[width=0.49\textwidth,trim=0 0 0 0,clip]{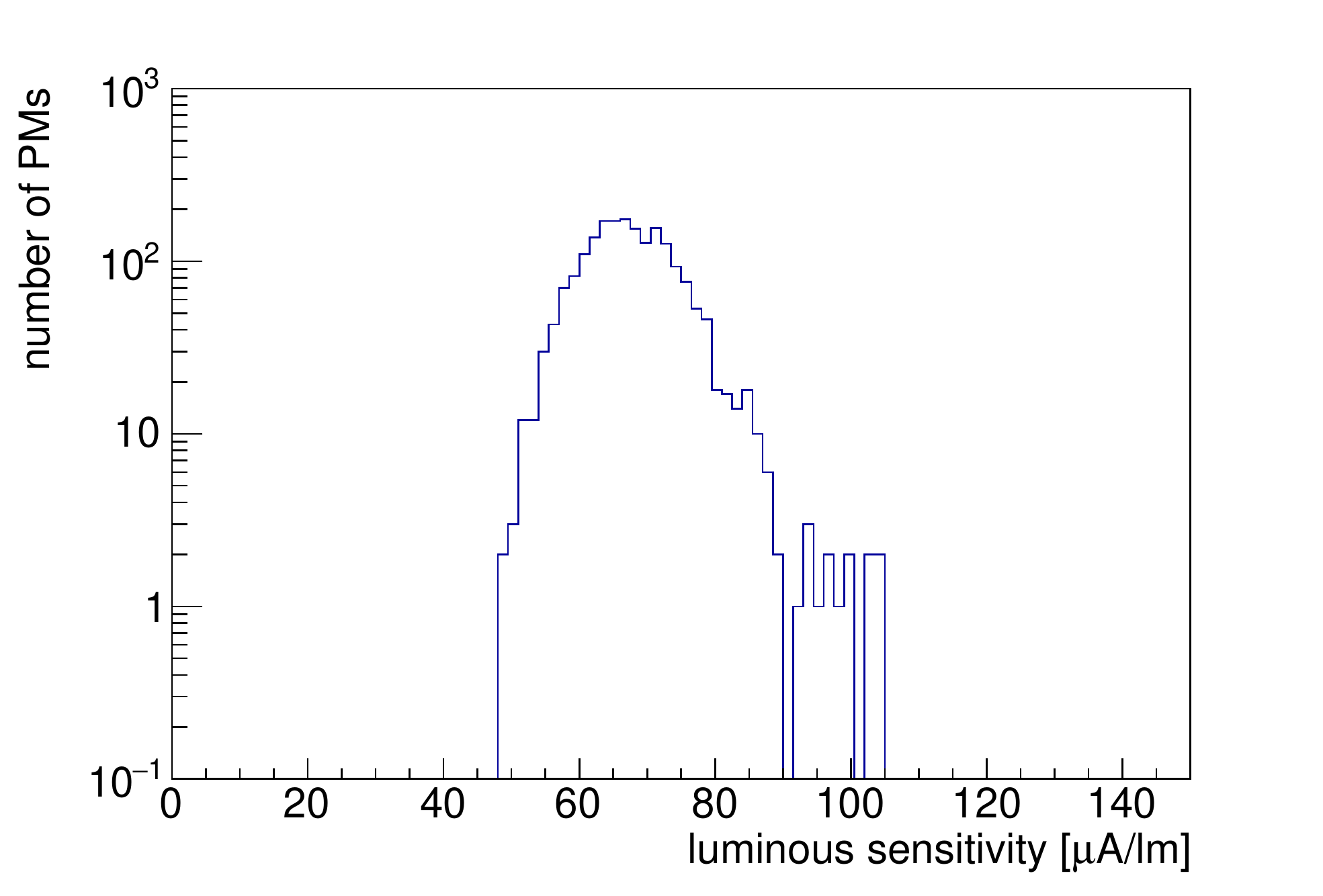}
% \put(-215,85){\colorbox{gray!0}{\makebox(5.5,45){}}} 
% \put(-210,75){\scriptsize \rotatebox{90}{Number of PMs}}
% \put(-100,3){\colorbox{gray!0}{\makebox(75.5,2){}}}
% \put(-120,-1){\scriptsize \rotatebox{0}{luminous sensitivity [$\mu$A / lm]}}
\includegraphics[width=0.49\textwidth,trim=0 0 0 0,clip]{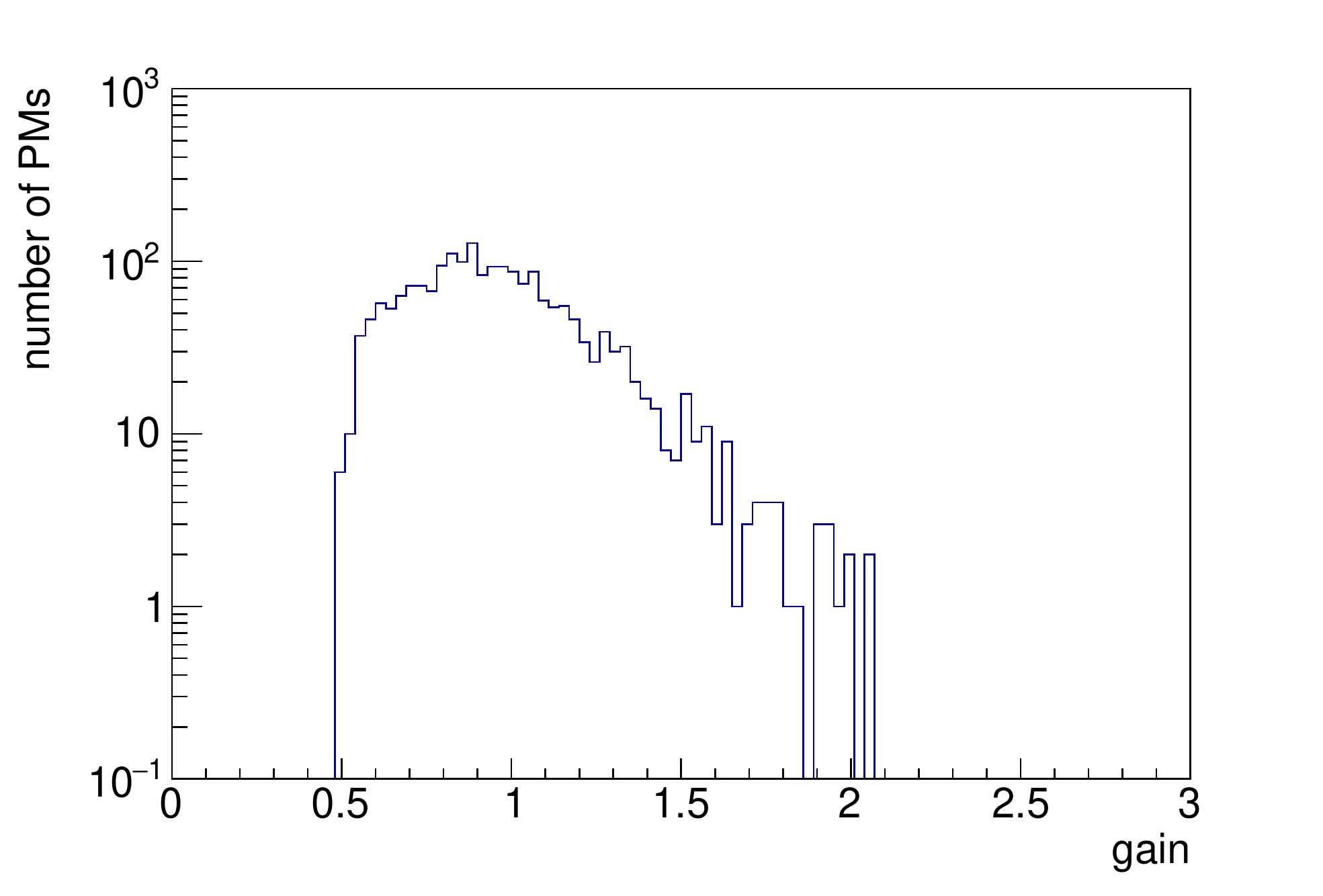}
%  \put(-215,85){\colorbox{gray!0}{\makebox(5.5,45){}}}
% \put(-210,75){\scriptsize \rotatebox{90}{Number of PMs}}
% \put(-50,3){\colorbox{gray!0}{\makebox(25.5,3){}}}
% \put(-40,-1){\scriptsize \rotatebox{0}{gain}}
\put(-15,13){\scriptsize \rotatebox{0}{$\times$ $10^6$}}
\caption{\label{fig:datasheet} 
%Characterization 
Manufacturer data for the 
%1952 installed 
R7400U-03 \pmts. Left: luminous sensitivity. Right: gain at 800~V.
A minimum gain of $0.5\times10^6$ was requested from the manufacturer.
}
\end{figure}

	The cathode luminous sensitivity quantity is proportional to the photo-current produced by an incident flux from a tungsten lamp operated at 2856 K. In figure~\ref{fig:datasheet}-left the cathode luminous
sensitivity for the installed PMs is presented, showing that the sample has an average of 68 $\mu$A lm$^{-1}$ with few devices reaching 100 $\mu$A lm$^{-1}$. All PMs underwent a laser acceptance test before 
% being installed 
installation
in the detector.

	The typical gain of the the R7400U-03 is 
%$0.5\times10^6$ 
$0.7\times10^6$ 
at 800 V and $1.5\times10^6$ at 900 V. The corresponding average charge generated per photo-electron is 
%80 fC 
110 fC 
and 240 fC, respectively. The frontend electronics is fully efficient only for charge 
% greater
larger 
than 100 fC (see section~\ref{subsec:nino}), as a consequence an operating voltage of 900 V was chosen. 
Since the maximum operating voltage for R7400U-03 is 1000 V, 
there is 
%plenty of room to further increase 
a possibility to increase
the applied voltage in case of ageing. For the NA62 RICH, PMs with a gain greater than $0.5\times10^6$ at 800 V 
% have been 
were
requested from the manufacturer (see figure~\ref{fig:datasheet}-right).

\begin{figure}
\centering
\includegraphics[width=0.6\textwidth]{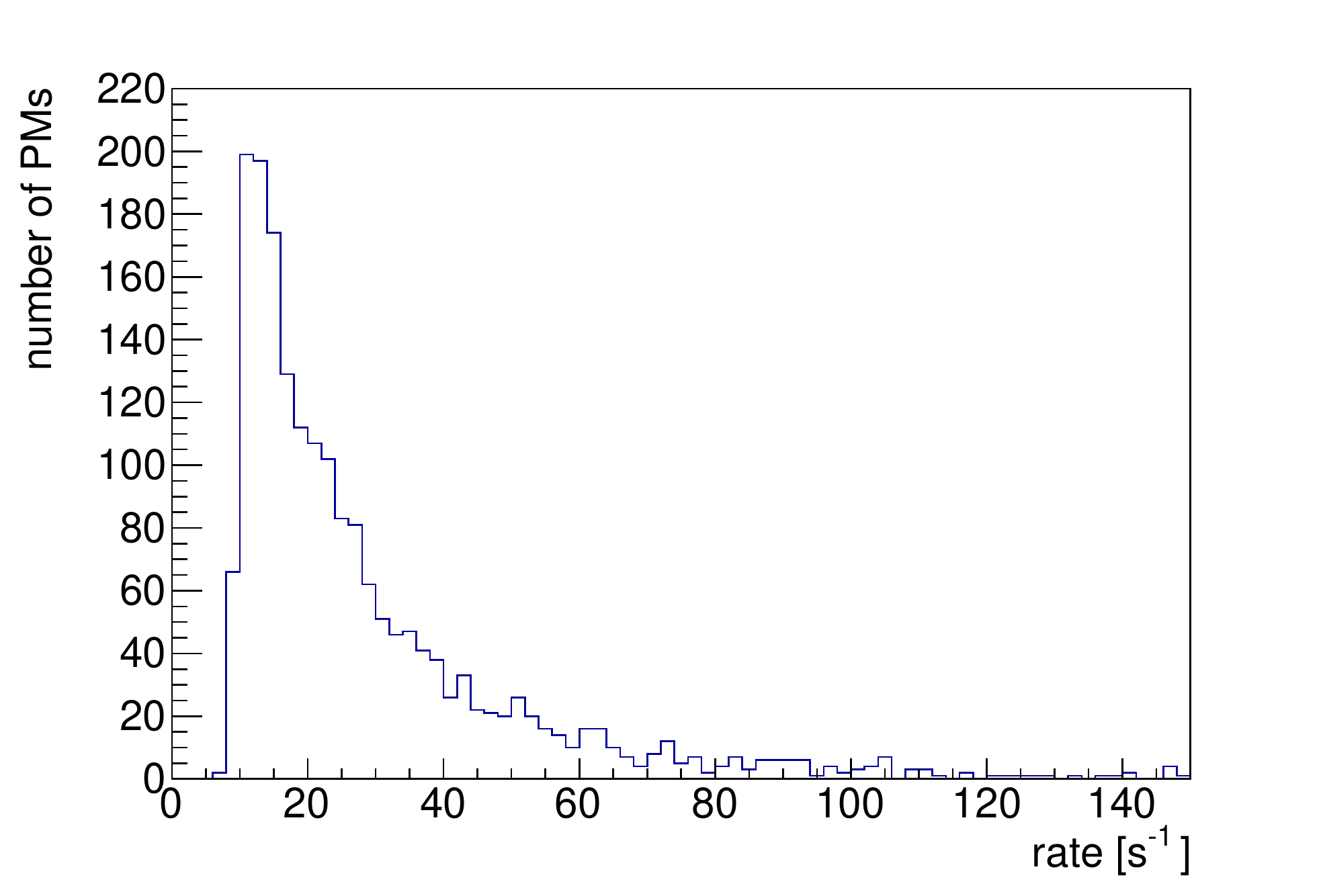}
%\put(-56.5,158){\colorbox{gray!0}{\makebox(45,1){}}}
% \put(-260,115){\colorbox{gray!0}{\makebox(5.5,45){}}}
% \put(-255,100){\scriptsize \rotatebox{90}{Number of PMs}}
% \put(-60,3){\colorbox{gray!0}{\makebox(30.5,3){}}}
% \put(-60,-1){\scriptsize \rotatebox{0}{rate [s$^{-1}$]}}
\caption{PM counting rate 
distribution
obtained with 
the beam off.} 
\label{fig:dark_distr}
\end{figure}

	The dark 
count 
rate of each individual PM is evaluated in situ 
% using the data acquisition 
from the PM counting rate 
when the beam is off. 
The measurement is performed periodically and is part of the 
characterization routine for 
detector maintenance and the response equalization. Figure~\ref{fig:dark_distr} shows the dark count rate measurements. The 98\% of the 1952 PMs have a dark count rate below 150 s$^{-1}$ with an average of 30 s$^{-1}$ and a typical value of 10 s$^{-1}$.

\subsection{Voltage divider}
%% \subsection{HV dividers}
\label{subsec:hvdividerl}
%%%%%%%%%%%%%%%%%%
A custom voltage divider 
% has been 
was
designed to replace the E5780 voltage divider (standard for R7400U-03) to overcome the problem of heat dissipation and to avoid the use of a cooling system for the PM disks. This precaution allows to prevent variation of the refractive index of the radiator due to temperature gradient along the vessel. 
The divider is encapsulated in a socket with the same geometrical characteristics and pin assignment as those of the E5780. Optical tightness is guaranteed by coating the circuit with a flame retardant, halogen free, low smoke and heat conductive resin. At 900 V bias, the custom divider has a current flow of 32 mA, to be compared with 320~mA of the E5780. The socket is equipped with two cables:  a 22 AWG twisted pair to feed the high voltage and an RG-178 coaxial cable to route the anode signal towards the frontend electronics.

%%%%%%%%%%%%%%%%%%
\subsection{High voltage power supply}
%% \subsection{HV power supply}
\label{subsec:hvpower}
%%%%%%%%%%%%%%%%%%
A reasonable compromise between the channel independence and the cost of the high voltage boards led to a distribution scheme in which a single HV channel feeds four PMs (referred as a cell). For the HV system two 
% kinds of HV 
different
boards are used, 12 modules of type CAEN A1733N
(12 channels) and 16 modules of type CAEN A1535SN (24 channels), for a total number of 528 output channels. 
% During physics runs all the cells have been operated at 900 V. In this condition
At 900 V
the absorbed current is about 130-140 $\mu$A for each cell, giving a total consumption of 
about
60~W.

%%%%%%%%%%%%%%%%%%
%%%%%%%%%%%%%%%%%%
\section{PM mechanics}
\label{sec:install}
%%%%%%%%%%%%%%%%%%
%%%%%%%%%%%%%%%%%%
To simplify access and avoid discharges,
the 
% photo-multipliers 
PMs
are mounted outside the radiator gas volume. 
A test 
%was 
performed during the R\&D phase 
%showing 
showed
that the selected PMs operating                                
in helium 
% start
started 
to produce discharges with supplied voltages greater than 600 V. 
The assembly consists
of two independent aluminium flanges: 
a 35 mm thick flange holding the PMs (PM disk)
%% (figure~\ref{fig:pmylodgingdisk})
and a 23 mm thick radiator flange.
% with quartz windows.  
%% (figure~\ref{fig:winstoconesdisk}).
%%%%%%%%%%%%%%%%%%
%\subsection{PMs lodging and closing disks}
%\subsection{Radiator flange and PM disk}
\subsection{Radiator flange}
\label{subsec:lodging disk}
\begin{figure}[htbp]
  \centering % \begin{center}/\end{center} takes some additional vertical space
  \includegraphics[width=0.49\textwidth,trim=0 0 0 0,clip]{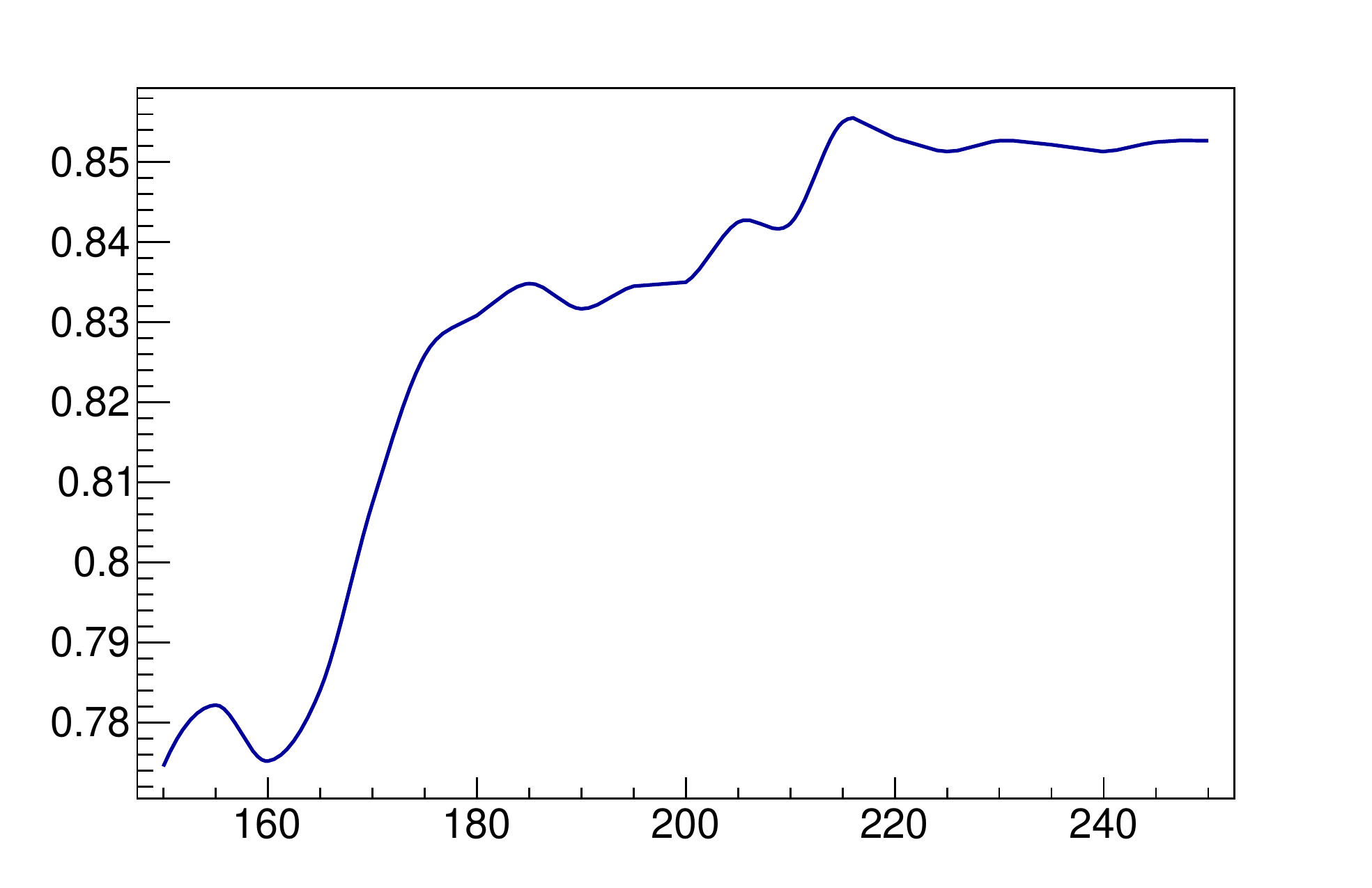}
      \put(-75,-1){\scriptsize wavelength [nm]}
      \put(-215,85){\scriptsize \rotatebox{90}{reflectivity}}
  \includegraphics[width=0.49\textwidth,trim=0 0 0 0,clip]{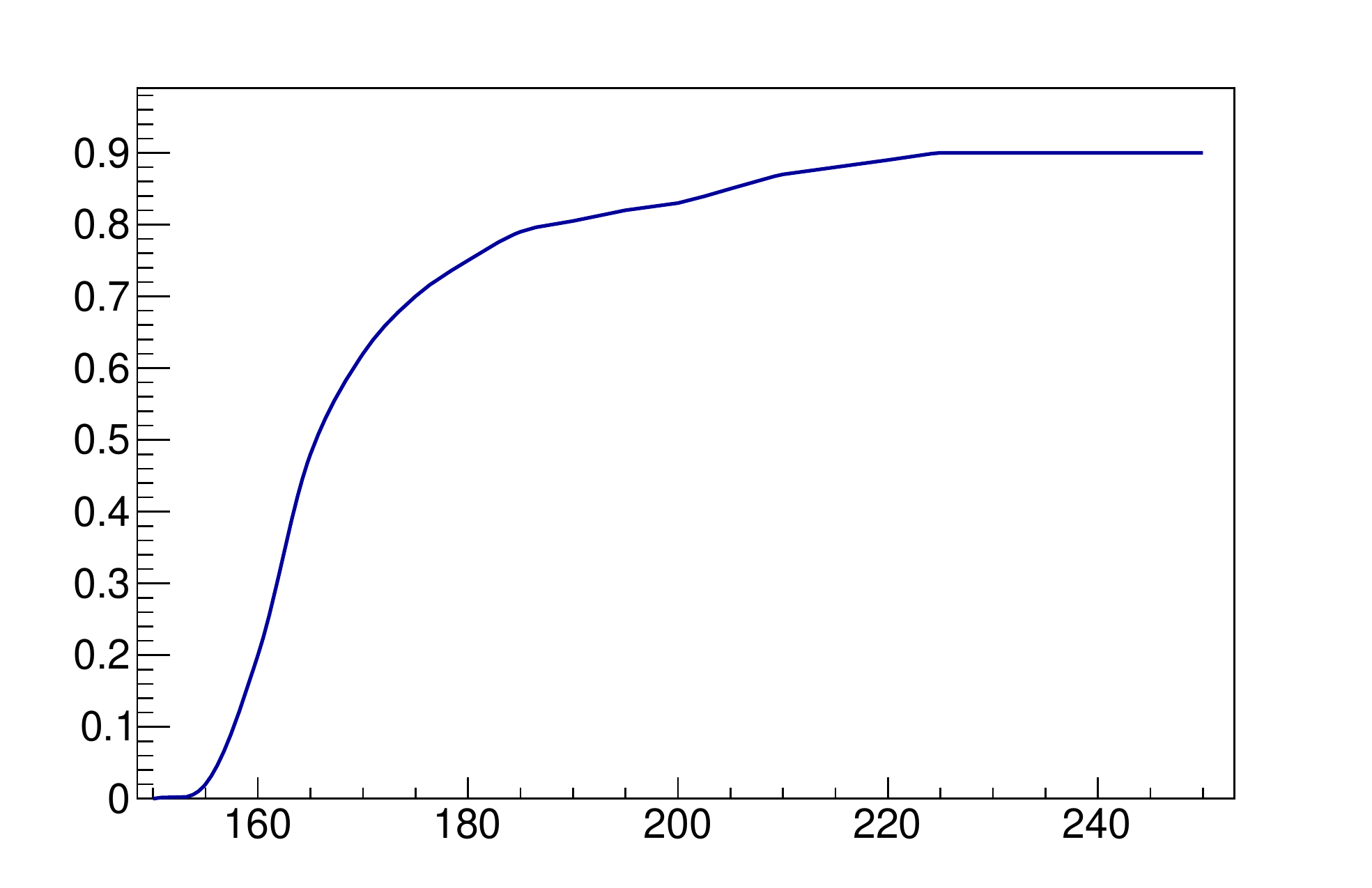}
      \put(-75,-1){\scriptsize wavelength [nm]}
      \put(-212,80){\scriptsize \rotatebox{90}{transmission}}
  \caption{\label{fig:reflectivitytransmission} 
  Left: mylar reflectivity. Right: quartz window transmission. 
}
\end{figure}

%%%%%%%%%%%%%%%%%%
The light entrance holes on the radiator flange have 
% the shape of a truncated cone 
a truncated cone shape,
approximating a
paraboloid ("Winston Cone"~\cite{winston}).
A Winston cone is designed to maximize the collection of incoming light by allowing off-axis rays to enter the exit aperture after multiple reflections. The optimized dimensions of the cone are 21.5 mm high, 18 mm wide at the entrance and 7.5 mm at the exit.
Each cone is covered with a highly reflective aluminized Polyethylenterephthalat
(Mylar\texttrademark) foil to funnel the light through the window aperture. 
The average reflectivity of the mylar foil 
measured for different samples as a function of the radiation wavelength in shown in figure~\ref{fig:reflectivitytransmission}-left.
%figure~\ref{fig:mylarreflectivityaverage} . 
%glued to the inner side of the Winston cone was measured 
%for different samples as a function of radiation length, see figure~\ref{fig:mylarreflectivity}
%and and their average in figure~\ref{fig:mylarreflectivityaverage}.

Outside each Winston cone, on the PM side, there is a 1.5 mm deep, 14 mm wide cylindrical hole to accommodate the quartz windows. Quartz windows, used to separate PMs from the neon gas,  are 12.7 mm wide, 1 mm thick and made of synthetic UV-grade fused silica. The quartz transmission is shown in  figure~\ref{fig:reflectivitytransmission}-right.
% figure~\ref{fig:fusedsilica}. 
The refractive index is 1.46 and the density is 2.20 g/cm$^3$. The windows are manufactured by Präzisions Glas \& Optik GmbH. 

%%%%%%%%%%%%%%%%%%
\subsection{PM disk}
\label{subsec:pmdisk}
%%%%%%%%%%%%%%%%%%
The PMs are mounted on the PM disk in front of the quartz windows. A cylindrical hole, 16.4 mm wide and 12.5 mm high, has been drilled in the disk for each PM, followed by a 17.5 mm wide and 20 mm high hole for the HV divider. A 1 mm thick O-ring (17.5 mm outer and 13.5 mm inner diameter, retained by a 1.5 mm thick groove in the hole, 1 mm above the end) has been placed in front of the PM and pressed against the quartz window to avoid the penetration of external light. A 5 mm thick O-ring (with the same outer and inner diameter as the 1 mm O-ring) has been placed on the back of the PM, after the end of the HV divider, to close the hole and avoid external light. 
This 
% latter 
O-ring also guarantees good thermal contact between the PM and the PM disk.

% \begin{figure}[htbp]
% \centering 
% \includegraphics[width=0.8\textwidth,trim=2 2 2 2,clip,angle=0]{MylarReflectivity}
% \put(-115,10){\colorbox{gray!0}{\makebox(15,10){wavelength [nm]}}}
% \put(-215,10){\colorbox{gray!0}{\makebox(60,10){}}}
% \put(-245,195){\colorbox{gray!0}{\makebox(100,10){}}}
% \put(-345,100){\rotatebox{90}{\colorbox{gray!0}{\makebox(70,10){Reflectivity [\%]}}}}
% \put(-334,70){\rotatebox{90}{\colorbox{gray!0}{\makebox(50,4){}}}}
% \put(-50,100){\colorbox{gray!0}{\makebox(45,15){}}}
% \caption{\label{fig:mylarreflectivityaverage} Mylar reflectivity.}
% \end{figure}

%%%%%%%%%%%%%%%%%%
%%\subsection{Quartz Windows}
%%\label{subsec:quartz}
%%%%%%%%%%%%%%%%%%
% \begin{figure}[htbp]
% \centering 
% \includegraphics[width=0.8\textwidth,trim=2 2 2 2,clip,angle=0]{QuartzTransmission}
% \put(-115,10){\colorbox{gray!0}{\makebox(15,10){wavelength [nm]}}}
% \put(-215,10){\colorbox{gray!0}{\makebox(60,10){}}}
% \put(-245,195){\colorbox{gray!0}{\makebox(100,10){}}}
% \put(-345,100){\rotatebox{90}{\colorbox{gray!0}{\makebox(70,10){Transmission [\%]}}}}
% \put(-334,70){\rotatebox{90}{\colorbox{gray!0}{\makebox(50,4){}}}}
% \put(-50,100){\colorbox{gray!0}{\makebox(45,15){}}}
% \caption{\label{fig:fusedsilica} 
%% Left: Quartz Window Transmission (nominal).
% Quartz window transmission. 
%(measured) for two samples.
% }
% \end{figure}

%%%%%%%%%%%%%%%%%%
%%%%%%%%%%%%%%%%%%
\section{Frontend electronics}
\label{sec:fee}
%%%%%%%%%%%%%%%%%%
%%%%%%%%%%%%%%%%%%
The frontend (FE) electronics is placed near the PM disks and is comprised of 64 boards housed in four crates. 
The design of custom FE boards is based on the NINO chip~\cite{nino}.

%%%%%%%%%%%%%%%%%%
\subsection{NINO chip}
\label{subsec:nino}
%%%%%%%%%%%%%%%%%%
The NINO chip is an 8-channel  ASIC 
%(application specific integrated circuit) 
developed at CERN for the ALICE experiment for precision time measurement.
% of the time-of-flight detector.
Each channel features an ultra fast low noise input amplifier followed by a discriminator stage.
% to produce extremely fast logic pulses when the injected charge exceeds a certain threshold.
The %ASIC
NINO chip 
has been adopted for the RICH %because of five main 
due to the following
characteristics:
input signal range between $100$ fC and $2$ pC (compatible with the \pmt output);
%NINO offers a very 
small time jitter ($<25$ ps for charge greater than $200$ fC);
%With 
$1$ ns peaking time (to sustain the RICH single channel rate up to few hundred kHz);
%The 
output 
LVDS
pulse width 
%is 
proportional to the input charge preserving charge information after digitization. 
%Finally 
The discrimination threshold can be set as low as $10$ fC %which is 
corresponding to
a small fraction of a typical \pmt signal. %response spectrum.

In addition the NINO has a fully differential circuit design that gives 
% large 
immunity against noise and %offers 
provides
a digital OR output used for trigger purposes. % the readout.
%Using the high voltage naming scheme, 
Each NINO chip serves two 4-channel HV cells,  
the 8-channel OR ouput is defined as a supercell.
The total number of supercells is 244.

%%%%%%%%%%%%%%%%%%
\subsection{Frontend boards}
\label{subsec:feeboard}
%%%%%%%%%%%%%%%%%%
\begin{figure}
\centering
\includegraphics[width=0.8\textwidth,trim=0 0 0 0,clip,angle=0]{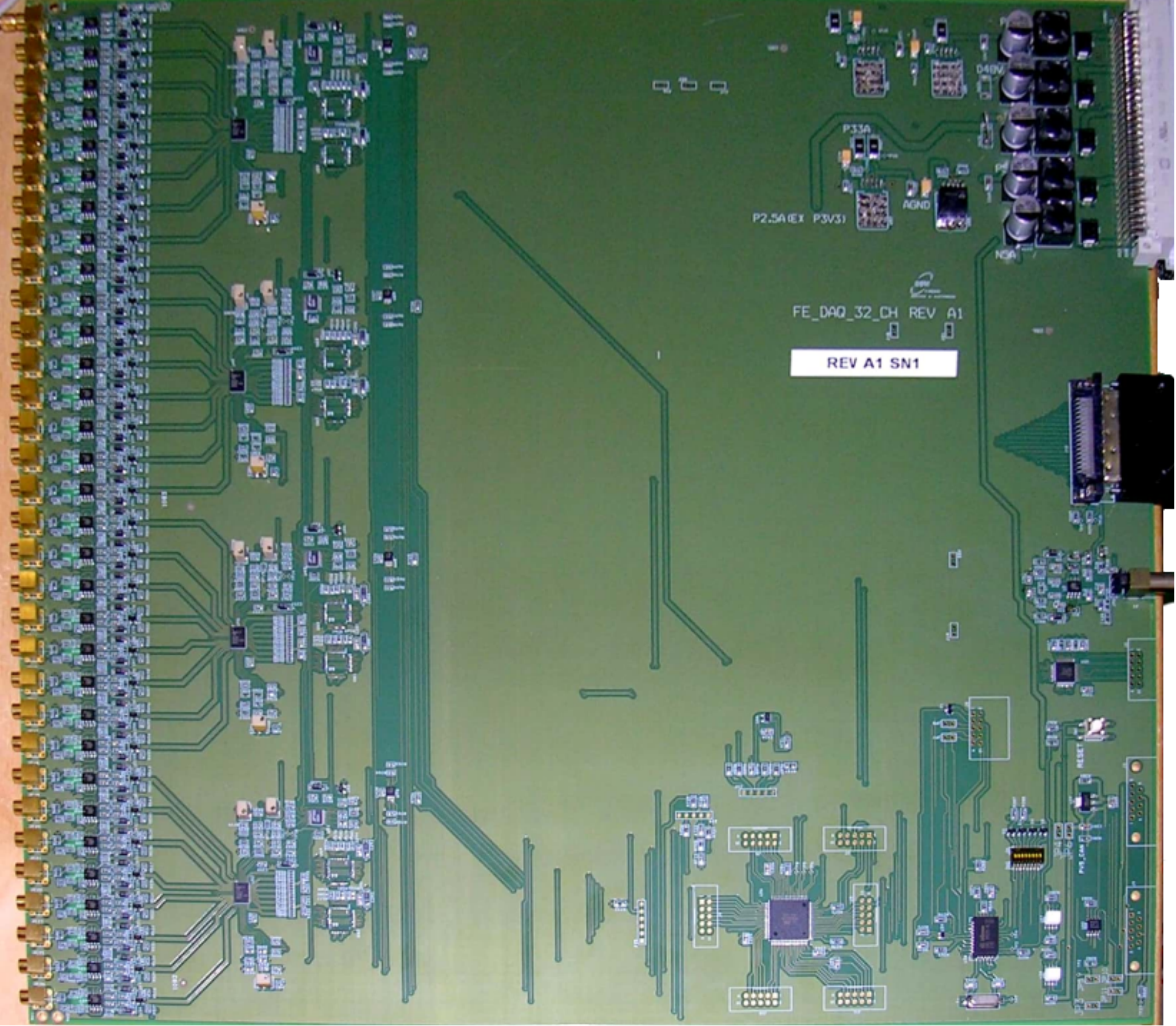}
\caption{The 32-channel custom frontend board based on the NINO chip.}
\label{fig:feb}
\end{figure}

Each FE board has 32 channels and houses four NINO chips. Each channel has a custom designed preamplifier (PA) with gain $\sim$3 and differential signal output. Special attention has been devoted to impedance adjustment: in order to avoid reflections each channel has a 50 $\Omega$ impedance both at the PA and NINO level. In addition to the ancillary components of NINO and PA, the frontend board houses the following parts: a power section with appropriate voltage regulators, a circuit for external test pulse distribution and a group of lines to set and monitor NINO parameters (threshold, pulse width and hysteresis). There are two options to set NINO parameters: a trimmer resistor or a DAC (digital to analog converter) remotely controlled via CAN bus. 

%The output signal is an LVDS pulse. 
The boards are powered by AGILENT N6700
mainframes. Four modules are installed in each mainframe to provide independent voltage lines: $+5$ and $-5$ V for the PA's, 
$+2.5$ V for the NINOs and a $+5$ V for the other digital logic circuits.
The total power consumption for each 
% sector 
crate
is 170 W.

%%%%%%%%%%%%%%%%%%
%%%%%%%%%%%%%%%%%%
\section{DAQ system}
\label{sec:daq}
%%%%%%%%%%%%%%%%%%
%%%%%%%%%%%%%%%%%%
The readout of the RICH
% , as most of other detector of NA62, 
is based on the NA62 common system composed of a mother board, the TEL62~\cite{tel62}
%, housing 
and
four daughter  
Time to Digital Converter Boards
(TDCB)~\cite{tdcb}. 
In total 
%the entire RICH is read by 
there are
five TEL62 boards, 
four boards to read the PM signals and one to read the signals from supercells.

%%%%%%%%%%%%%%%%%%
\subsection{The TDCB}
\label{subsec:tdc}
%%%%%%%%%%%%%%%%%%
The TDCB 
% is designed to receive in input a maximum of 
input is designed to receive up to
128 channels (4 connectors with 32 channels each). The requirements of a good time resolution led to the choice of the CERN high-performance time-to-digital converter (HPTDC)~\cite{hptdc}. The HPTDC output provides the leading and trailing time of the LVDS input signal with 100 ps least significant bit (LSB). Each TDCB houses 4 HPTDC and 1 FPGA used to transfer data from each single HPTDC to the TEL62 board. The FPGA is also used to configure the TDCB. 

%%%%%%%%%%%%%%%%%%
\subsection{TEL62}
\label{subsec:tel62}
%%%%%%%%%%%%%%%%%%
The TEL62 is a general-purpose data acquisition board, an updated version of the TELL1 board designed by the EPFL for the LHCb experiment at CERN~\cite{tell1}. The board 
% comprises 
hosts
five FPGA.
Four of them are called PP (Pre-Processing), 
each one handling the data coming from a TDCB.
The fifth FPGA, called SL (Sync-Link), is used
to merge the data flux of the four PPs. The SL also sends the data to the online PC farm via a 4 gigabit ethernet daughter card. The TEL62 can also produce trigger primitives and send them to a central L0 trigger processor~\cite{l0system},
% The
whose 
expected maximum output rate is 1 MHz.

%%%%%%%%%%%%%%%%%%
\subsection{RICH in L0 Trigger}
\label{subsec:trigger}
%%%%%%%%%%%%%%%%%%
The RICH is one of the detectors used in the L0 Trigger system of NA62, producing the reference time for most trigger configurations. The RICH primitives are generated by the TEL62 dedicated to the readout of supercell signals. The trigger algorithm is based on the clustering of supercell signals and on the computation of cluster multiplicity and average time. In the PP, if at least two hits have a time difference 
% lower 
less
than 6.25 ns, a sub-cluster primitive is generated. The SL firmware stores the sub-clusters coming from different PPs into one buffer. If two or more sub-clusters are closer in time than 6.25 ns, the sub-cluster multiplicities are summed and define the cluster multiplicity; the cluster time is computed as the weighted average of sub-cluster times. 
The RICH primitive is generated if the cluster multiplicity is greater or equal to an adjustable threshold
% .
that allows to keep the L0 primitive rate within the design limit of 10 MHz. 
The threshold was 
initially
set to 3 
%during the 2016 run 
and 
lowered
to 2 
% starting from the 
in
2017 
% run.
after 
% the
an 
L0 system improvement.

%%%%%%%%%%%%%%%%%%
%%%%%%%%%%%%%%%%%%
\section{Light detection efficiency}
\label{sec:efficiency}
%%%%%%%%%%%%%%%%%%
%%%%%%%%%%%%%%%%%%

The relative  efficiency of Cherenkov photon detection was measured for the 2016-2018 data taking period for most
of the RICH PMs. Here the results obtained for the year 2017 are shown.

With a tight selection of $K\rightarrow \pi^0 e^+ \nu$ decays (Ke3) a clean positron sample has been obtained without using
the RICH information. 
% added
For positrons the average ring radius and number of hits are constant in the momentum range exploited in the experiment.
The background from other decays is at the level of $10^{-3}$.
The Cherenkov photons emitted by positrons are requested to be in the geometric acceptance of the RICH 
% (impinging on the mirror surface). 
mirrors.
This requirement together with the kinematic constraints derived by Ke3 decay selection does not allow to measure the
efficiency for all the PMs due to low statistics. 
%For  2 PMs in the Saleve flange and 39 in the Jura one the efficiency was not
%measured 
The efficiency was not measured for 2 PMs in the right disk (with respect to the beam direction) and 37 in the left one,
corresponding to $\sim2\%$ of PMs.

%%%%%%%%%%%%%%%%%%
%%%%%%%%%%%%%%%%%%
\subsection{RICH performance stability}
\label{subsec:stability}
\begin{figure}[h]
\centering
  \includegraphics[width=0.49\textwidth,trim=0 0 0 0,clip]{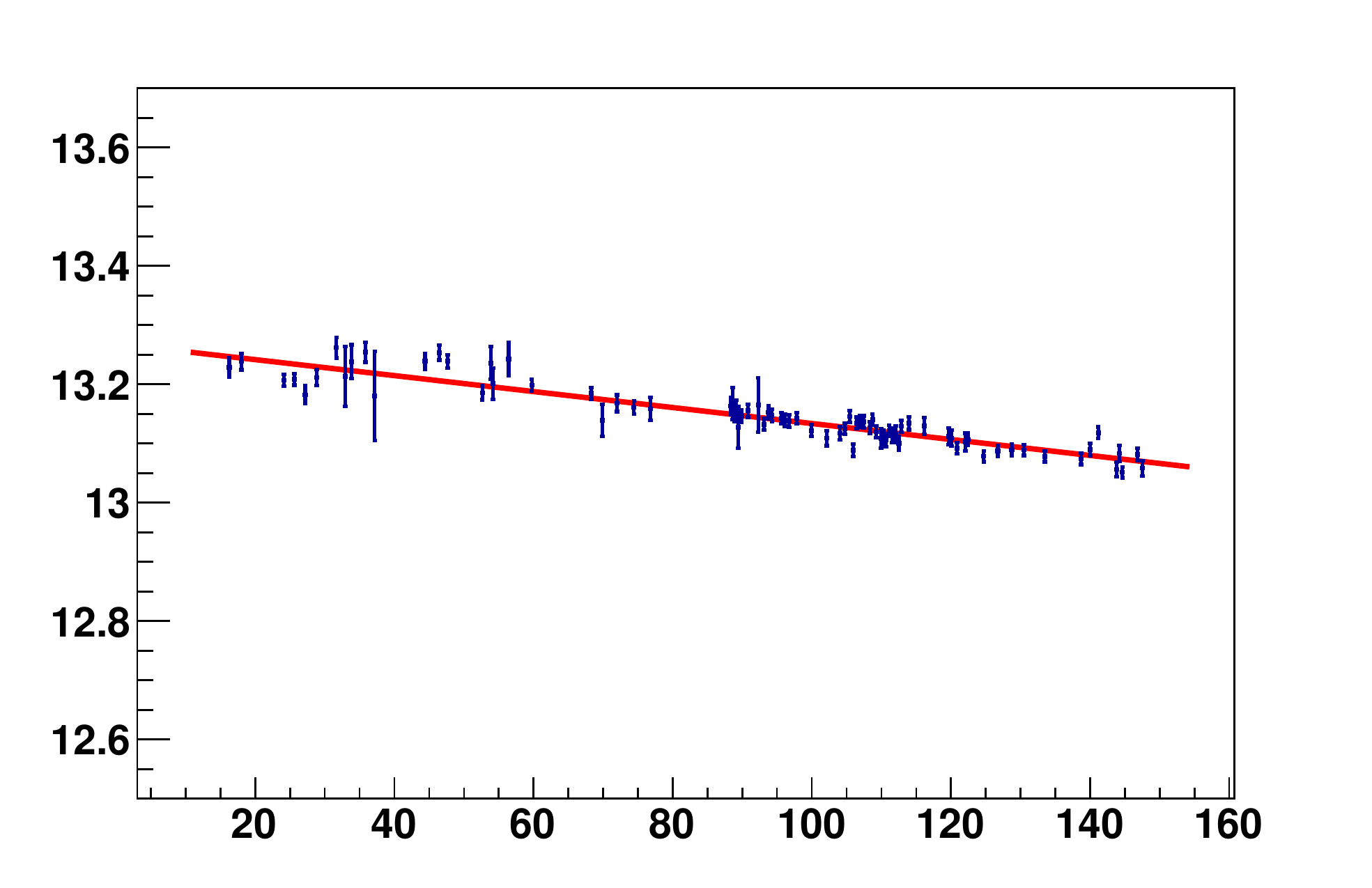}
        \put(-60,-1){\scriptsize time [days]}
        \put(-215,100){\scriptsize \rotatebox{90}{$\overline{N}_{hits}$}}
  \includegraphics[width=0.49\textwidth,trim=0 0 0 0,clip]{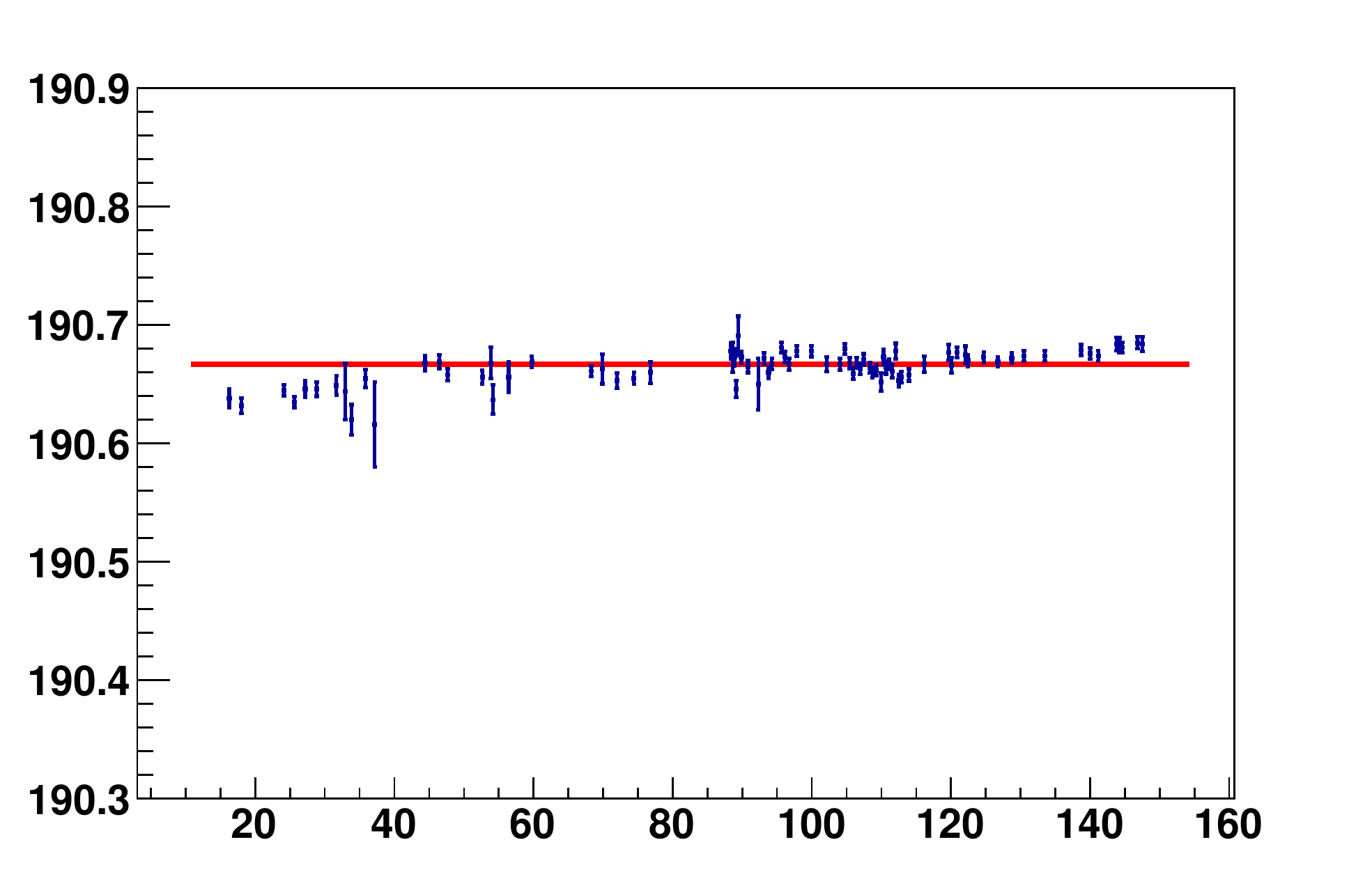}
        \put(-60,-1){\scriptsize time [days]}
        \put(-215,110){\scriptsize \rotatebox{90}{$\overline{R}$}}
\caption{Performance stability for positron rings. 
% Left: Nhits vs time. Right: ring radius vs time.
Left: $\overline{N}_{hits}$ vs time. Right: $\overline{R}$ vs time. 
Vertical bars represent statistical errors.
For the definition of $\overline{N}_{hits}$ and $\overline{R}$ see text.
}
\label{fig:stability}       % Give a unique label
\end{figure}

%%%%%%%%%%%%%%%%%%
%%%%%%%%%%%%%%%%%%
First, the stability of the RICH performance is checked. 
The number of hits and ring radius
% $N_{hits}$ and $R$ 
for positron rings 
% as a function of time for the 2017 data are shown in figure~\ref{fig:stability}.
are averaged over a certain time period (typically one day long) and the average values $\overline{N}_{hits}$ and $\overline{R}$ are plotted as a function of time (see figure~\ref{fig:stability}).

% The 
$\overline{N}_{hits}$ exhibits a slow decrease $\sim1.4 \cdot 10^{-3}$ hits/day, while $\overline{R}$ is stable within 0.04 mm (or $2\times 10^{-4}$ in relative terms). The average radius value is $R_e$ = 190.67 mm.

The obtained result proves the performance stability and allows to calculate the expected value of  $\overline{N}_{hits}$
and $\overline{R}$ for any time period.

%%%%%%%%%%%%%%%%%%
%%%%%%%%%%%%%%%%%%
\subsection{Light detection efficiency measurement}
\label{subsec:efficiency}
%%%%%%%%%%%%%%%%%%
%%%%%%%%%%%%%%%%%%
%First, 
For each selected Ke3 event the expected position of the ring center and the expected radius 
% $R_e$ 
of the $\beta=1$ Cherenkov ring of the positron at the PM flanges are
calculated exploiting the track direction from the spectrometer and the known value $R_e$.
% of the refractive index.
Defining $\Delta=D-R_{e}$ as the difference between $D$, the distance of the PM center
from the expected ring center, and the expected ring radius $R_{e}$, $| \Delta |$ is required to be
lower than 3 mm in order that the ring falls in the PM active area. 
For each PM satisfying such a condition
a "normalization" counter $n_{PM}$ is incremented by one.

%from the
%center of the PMs photocathode surface.

Afterwards, for the same event and for each PM included in the normalization subset with a hit
(without any assumption on the ring from the RICH reconstruction) in time with the
positron (the KTAG time of the event is used) an "efficiency" counter $e_{PM}$ is incremented by a weight $w$ defined as:
\begin{equation}
% w=(W_{Ref}*\sqrt(R^2.-D^2)-(1.-W_{Ref})*\sqrt(r^2.-D^2))/[W_{Ref}*(R-r)+r]
   w(\Delta) = \left[W_{Ref} \cdot R + (1 - W_{ref})\cdot r \right] / \left[W_{Ref} \cdot \sqrt{R^2 - \Delta^2}+(1-W_{Ref}) \cdot \sqrt{r^2-\Delta^2}\right]
\end{equation}
where $R=9$ mm is the radius of the entrance surface of the Winston Cone, $r=3.75$ mm is the radius
of the exit surface of the Winston cone (corresponding to the active surface of the PM photocathode),
$W_{Ref}=0.85$ is an estimation of the integrated reflectivity of the internal surface of the Winston
cone.
This weight $w(\Delta)$ is applied 
%to remove the dependence of the measured efficiency from the different illumination
%received by each PM from Cherenkov light.
to take into account  the dependence of the measured efficiency on the amount of Cherenkov light reaching a single PM.
A further correction $\alpha(\Delta)$ of the order of $1\%$ is applied to remove the residual $\Delta$ dependence  due to mirror misalignment,
thus $e_{PM} = \sum\nolimits_{i} w_i (\Delta) \alpha_i(\Delta)$ where $i$ is the event number.

%The efficiency of each PM is calculated as the ratio of the two counters 
The light detection efficiency related to a single PM is calculated as the ratio of two counters
$\epsilon_{PM}=e_{PM}/n_{PM}$
after all events in the Ke3 sample have been considered.
% The value $\epsilon_{PM}=e_{PM}/n_{PM}$, calculated using the $w$ values, 
% corresponds to the efficiency 
This value corresponds to the integrated efficiency
for a Cherenkov ring crossing the center of the PM
%.
and takes
% has no meaning on an event-by-event basis: given that the number
% of emitted photons 
% and their azimuthal distribution with respect to the particle direction
% is not known for a particular event, 
% the PM performance can be described only considering a large sample of events.
% $\epsilon_{PM}$ corresponds to an integrated efficiency taking 
into account all possible effects: the number of Cherenkov photons emitted by the positron passage 
in the radiator (determined by the figure of merit of the RICH detector), the reflectivity and the dispersion
introduced by the mirrors, the probability of absorption and diffusion in the radiator, the probability of photon transmission in
the quartz window, the QE of the PM and the readout efficiency of the RICH.

%Therefore is not possible to give to these measurements an absolute meaning, the QE 
%of the PM is only one factor entering in the product.
%Nevertheless the efficiency measurement can be used 
The light detection efficiency measurement is used
to compare different PMs, to identify misbehaving PMs and
to study the RICH response for different periods of the data taking, in order to spot possible ageing
or degradation effects.

\begin{figure}[h]
\centering
  \includegraphics[width=0.49\textwidth,trim=0 0 0 0,clip]{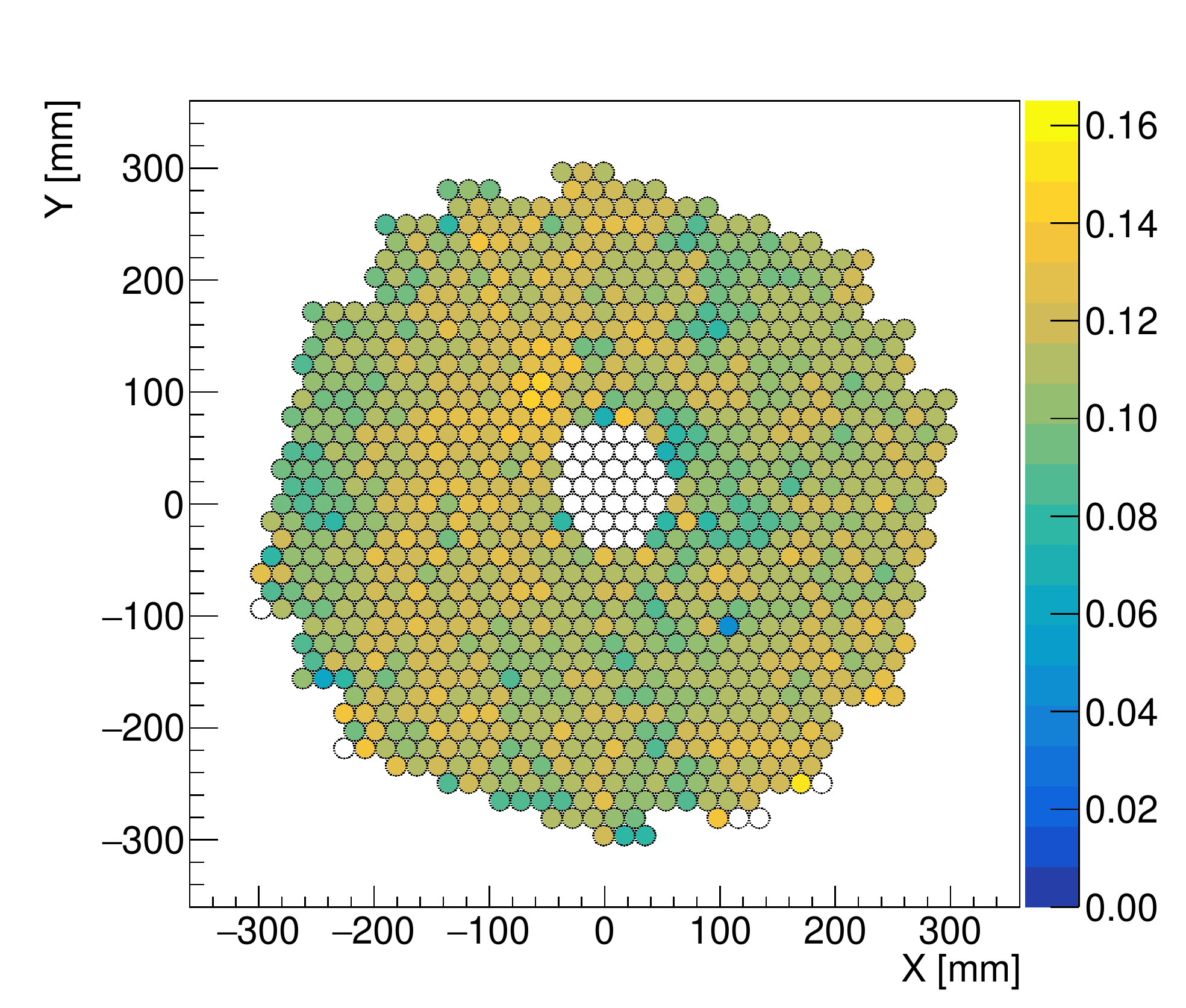}
  \includegraphics[width=0.49\textwidth,trim=0 0 0 0,clip]{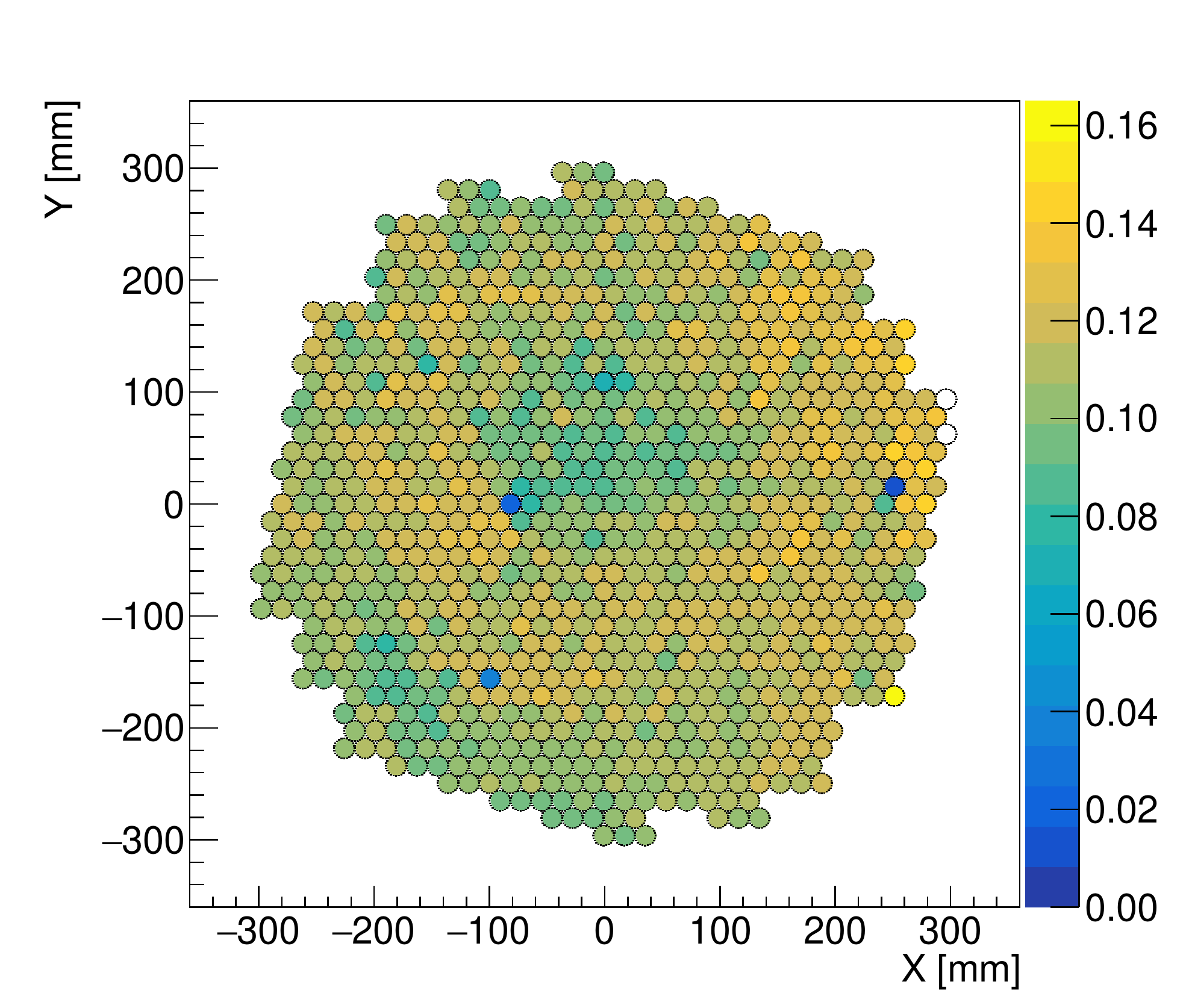}
\caption{The relative efficiency map for the 2017 data for the left (left) and right (right) PM disks. 
The disks orientation is defined  with respect to the beam direction looking from upstream.
The PMs marked in white correspond to those for which the measurement is not performed.}
\label{fig:flanges}       % Give a unique label
\end{figure}

\begin{figure}[h]
\centering
\includegraphics[width=0.6\textwidth]{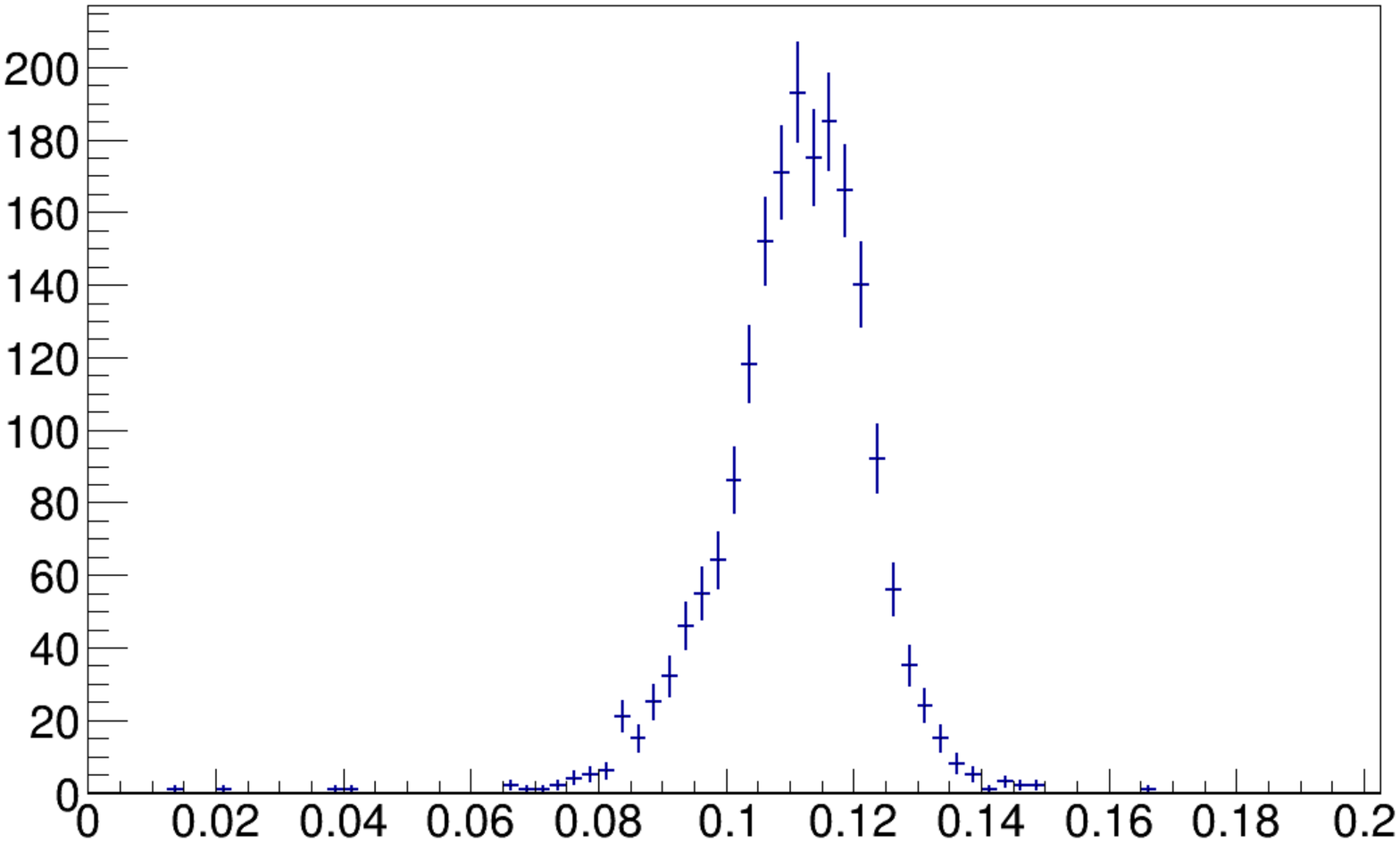}
        \put(-60,10){\scriptsize efficiency}
        \put(-255,100){\scriptsize \rotatebox{90}{ Number of PMs }}
\caption{
%Number of PMs as a function of the efficiency measured with the 2017 data.
Light detection efficiency distribution for 2017 data.
}
\label{fig:1Deff}       % Give a unique label
\end{figure}

%In figure~\ref{fig:flanges} the map of the relative efficiencies defined above and measured with the data collected in October 2017 
%is shown for the two flanges of the RICH.
Figure~\ref{fig:flanges} illustrates the relative efficiency map for two PM disks of the RICH.
%The efficiency distribution is also shown in figure~\ref{fig:1Deff} and an average efficiency of $9.9\%$ is measured.
The efficiency distribution is shown in figure~\ref{fig:1Deff}; the average efficiency is found to be $11.1\%$.
Four PMs ($\sim 2\text{\textperthousand}$ of the total sample) with problematic behaviour are identified.
%% This measurement matches the expectations derived by the RICH design. 
% and agrees with an independent measurement of average number of hits for positron rings.
%% The same measurement was performed on 2016 and 2018 data. 
%% For the 2016 data comparable results are obtained, in 2018 
%an average reduction of the relative efficiency of $2\%$ is measured. 
%% the average relative efficiency decreased by $2\%$.
%% The same relative decrease has been observed in the average number of hits for positron rings measured for the same data samples.
%% It is not possible to disentangle the source of such reduction, the main candidates are
%% the degradation of 
%the 
%% mirror reflectivity and the ageing of 
%the 
%% PMs.

%%%%%%%%%%%%%%%%%%
%%%%%%%%%%%%%%%%%%
\section{Time resolution}
\label{sec:timeresolution}
%%%%%%%%%%%%%%%%%%
%%%%%%%%%%%%%%%%%%
One of the main characteristics of the PM and readout performance is the time resolution. The measurement of the event and single hit time resolution is presented below.

%%%%%%%%%%%%%%%%%%
\subsection{Event time resolution}
\label{subsec:event}
\begin{figure}[htbp]
  \centering % \begin{center}/\end{center} takes some additional vertical space
  \includegraphics[width=0.49\textwidth,trim=0 0 0 0,clip]{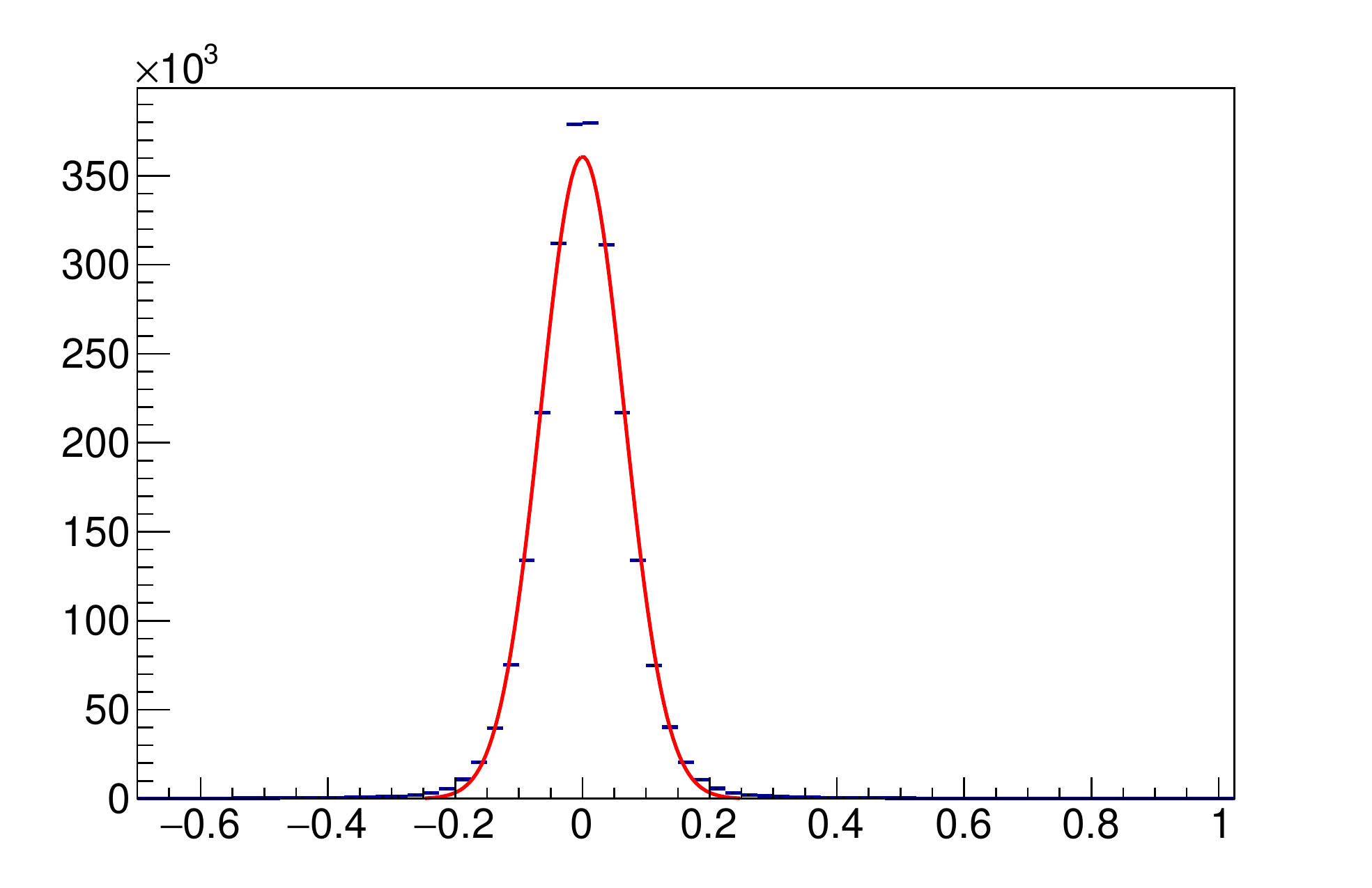}
      \put(-75,-1){\scriptsize 0.5 ($T_{1}$ - $T_{2}$) [ns]}
      \put(-210,65){\scriptsize \rotatebox{90}{events / (0.025 ns)}}
   % \includegraphics[width=0.49\textwidth,trim=0 0 0 0,clip]{event_time_resolution_wrt_cedar_2017A}
    %   \put(-102,-1){\scriptsize ($T_{RICH}$ - $T_{KTAG}$) [ns]}
    %  \put(-210,65){\scriptsize \rotatebox{90}{events / (0.025 ns)}}
   \caption{\label{fig:event_time_resolution} 
   % Left: 
   $\Delta T = 0.5\cdot(T_1 - T_2)$ distribution for 2017 data, the gaussian fit gives $\sigma (\Delta T) \simeq$ 0.07 ns.
   % Right:~$T_{RICH} ~- ~T_{KTAG}$ distribution for 2017 data, the gaussian fit gives $\sigma(T_{RICH}~-~T_{KTAG})~\simeq$~0.14 ns.
   For the definition of $T_1$ and $T_2$, see text.}
\end{figure}

%%%%%%%%%%%%%%%%%%
The event time resolution is measured using a clean sample of positron rings
% , since for positrons the average number of hits is constant in the momentum range exploited in the experiment. 
% This sample is created by selecting $K^+\to e^+ \nu_e \pi^0$ decays from the data collected in 2017. 
% The rings are required to be in the acceptance of the PM plane,
% while the light cones must be within the acceptance of mirrors and should not touch the beam pipe.
similar to the one described in Section~\ref{sec:efficiency} but with 
% tighter 
% cuts on the acceptance:
% the rings are required to be in the acceptance of the PM plane, while the light cones must be within the
% acceptance of mirrors and should not touch the beam pipe. 
a  tighter 
selection: the rings are required to be in the
acceptance of the PM plane, while the light cones 
should not touch the beam pipe.
These cuts are applied to remove rings that are
partially outside the PM acceptance: such rings are useful for the efficiency studies but could
bias the time resolution measurement due to lower number of hits per ring.
The details of the 
event
selection can be found in~\cite{alignment}.
In order to perform the time resolution measurement independently of other detectors, RICH hits associated with a ring are randomly split into two groups, 
the average time of each group $T_1$ and $T_2$ is calculated and  a new variable is introduced:
%$\Delta T = T_1 - T_2$. 
$\Delta T = 0.5 \cdot (T_1 - T_2)$. 
The event time is equal to $T_{RICH} \simeq 0.5\cdot(T_1 + T_2)$.
Assuming the hit independence, $\sigma(T_1 - T_2) \simeq \sigma(T_1 + T_2)$ and hence 
%$\sigma(T_{RICH}) \simeq 0.5 \cdot \sigma(\Delta T)$, 
$\sigma(T_{RICH}) \simeq \sigma(\Delta T)$, 
where $\sigma$ is the gaussian width. 
The distribution of the $\Delta T$ is shown in 
 figure~\ref{fig:event_time_resolution}.
 %-left. 
 The gaussian fit gives
$\sigma(\Delta T) \simeq 0.07$ ns. 
%and $\sigma(T_{RICH}) \simeq 0.07$ ns.
% The distribution of ($T_{RICH} - T_{KTAG}$) is shown in figure~\ref{fig:event_time_resolution}-right.
% The gaussian width is  $\sigma (T_{RICH} - T_{KTAG}) \simeq$0.14 ns which is larger
% than $\sigma (T_{RICH})$ due to the $T_{KTAG}$ resolution (0.07 ns~\cite{na62det}) and 
% contribution from systematic uncertainties of the 
% TDCB
% time offsets.
The wide fit range includes non-gaussian tails which results in a poor agreement between the data and fitting function in the central region.
The non-gaussian tails in the $\Delta T$ distribution are mostly due to the non-gaussian tails of the single-hit time distribution and 
to residual systematic effects of the TEL62 time offsets.

% \begin{figure}[h]
%   \begin{minipage}[t]{0.45\textwidth}
%      \centering
%      \includegraphics[width=6.0cm , angle=0]{event_time_resolution_dt_2017A}
%   \end{minipage}  
%   \hspace{0.2cm}
%   \begin{minipage}[t]{0.45\textwidth}
%       \centering
%       \includegraphics[width=6.0cm , angle=0]{event_time_resolution_wrt_cedar_2017A}
%    \end{minipage} 
%    \caption{Left: $\Delta T = 0.5\cdot(T_1 - T_2)$ distribution for 2017 data, the gaussian fit gives $\sigma (\Delta T) \simeq$ 0.07 ns.
%    Right:~$T_{RICH} ~- ~T_{KTAG}$ distribution for 2017 data, the gaussian fit gives $\sigma(T_{RICH}~-~T_{KTAG})~\simeq$~0.14 ns.
%    For the definition of $T_1$ and $T_2$, see text.}
%    \label{event_time_resolution}
 %   \begin{picture}(1,1)
%       \put(112,62){\scriptsize 0.5 ($T_{1}$ - $T_{2}$) [ns]}
%       \put(288,62){\scriptsize ($T_{RICH}$ - $T_{KTAG}$) [ns]}
%       \put(10,105){\scriptsize \rotatebox{90}{events / (0.025 ns)}}
%       \put(215,105){\scriptsize \rotatebox{90}{events / (0.025 ns)}}
%     \end{picture} 
% \end{figure}

%%%%%%%%%%%%%%%%%%
\subsection{Single hit time resolution}
\label{subsec:singlehit}
\begin{figure}[htbp]
  \centering % \begin{center}/\end{center} takes some additional vertical space
  \includegraphics[width=0.49\textwidth,trim=0 0 0 0,clip]{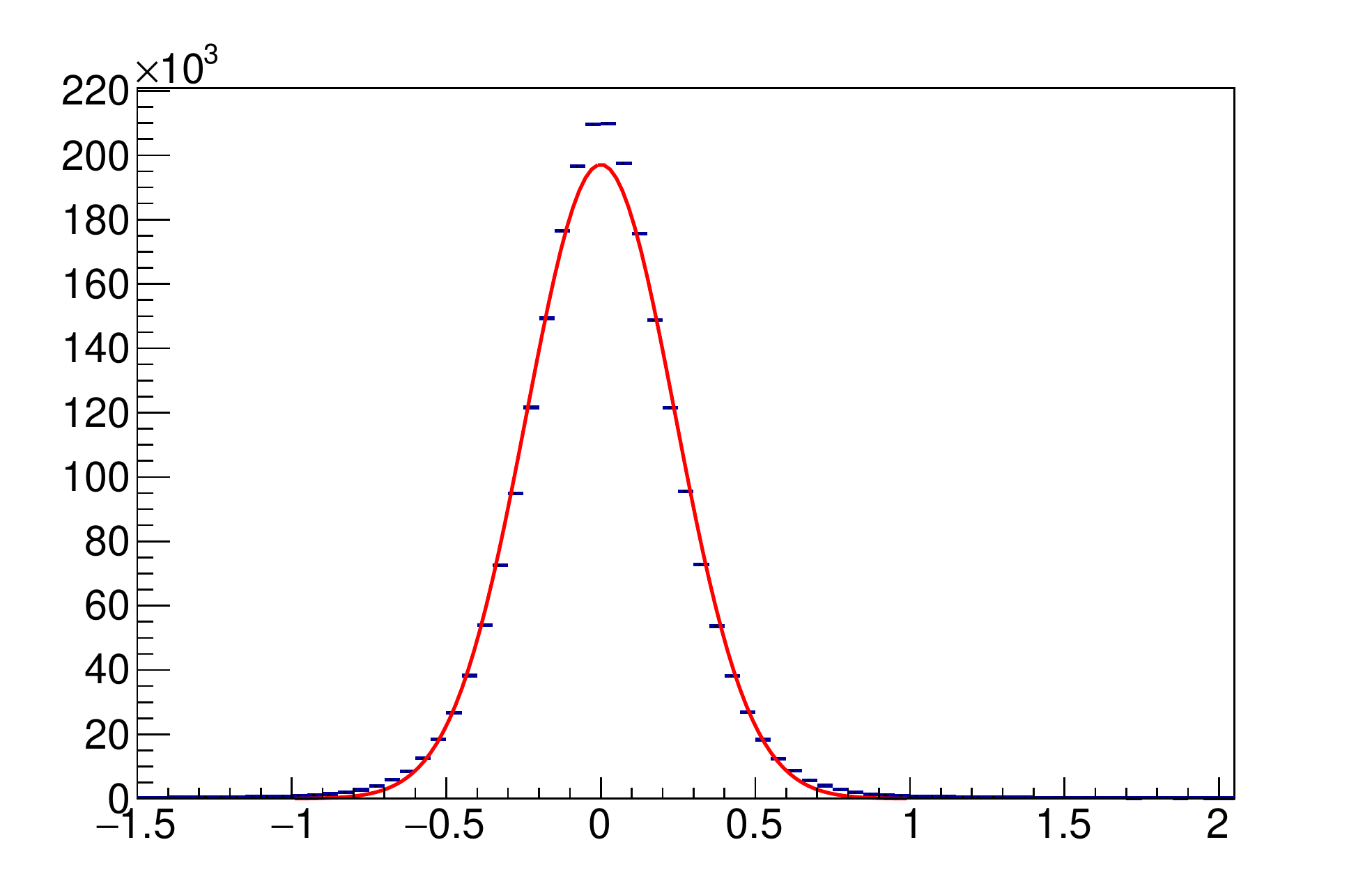}
      \put(-50,-1){\scriptsize Pull [ns]}
      \put(-210,65){\scriptsize \rotatebox{90}{events / (0.025 ns)}}
   \includegraphics[width=0.49\textwidth,trim=0 0 0 0,clip]{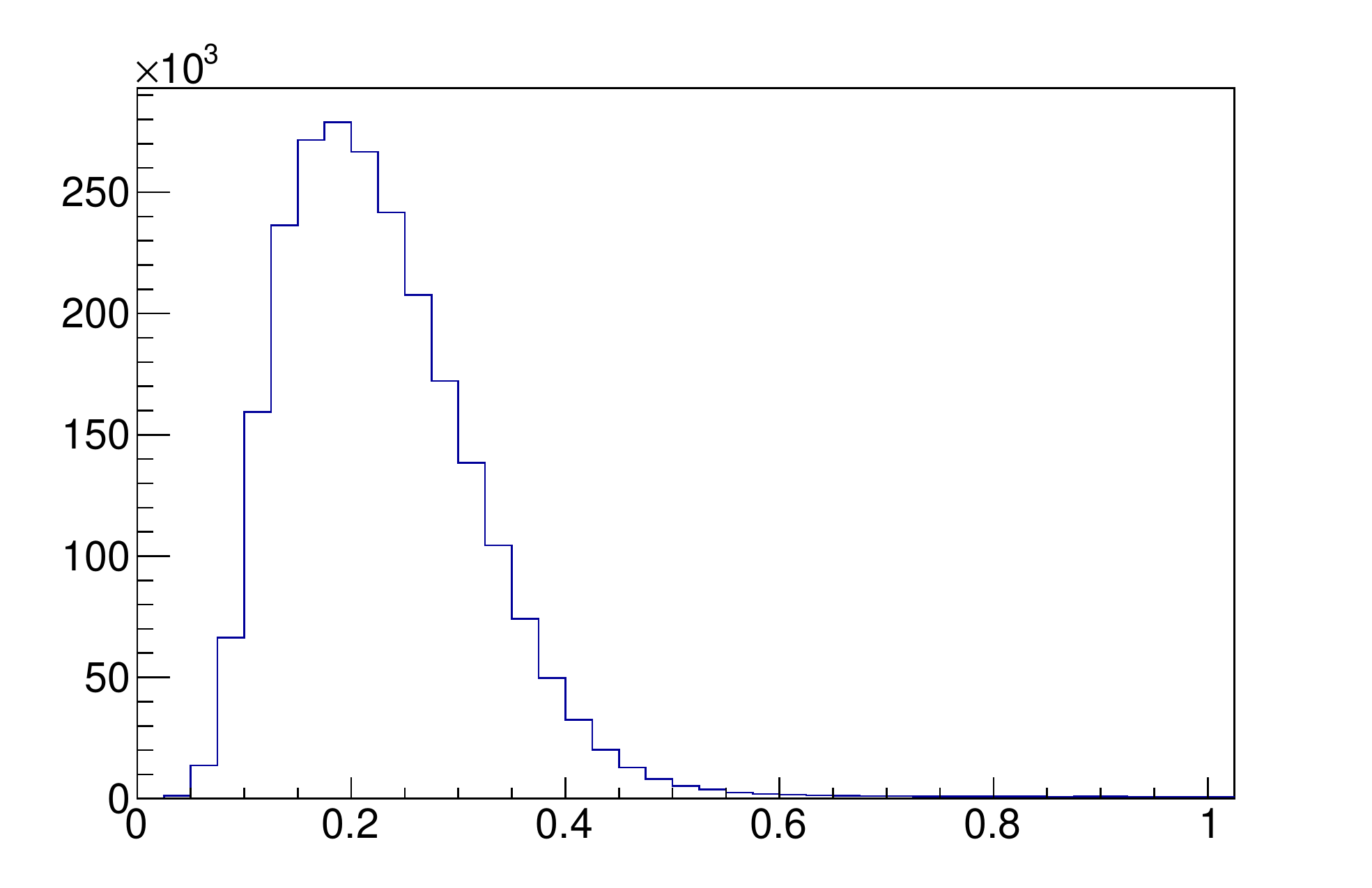}
      \put(-50,-1){\scriptsize $\sqrt{D}$ [ns]}
      \put(-210,65){\scriptsize \rotatebox{90}{events / (0.025 ns)}}
   \caption{\label{fig:single_hit_time_resolution} 
   Left: $Pull$ distribution for 2017 data, the gaussian fit gives $\sigma$$\simeq$0.24 ns.
  Right: $\sqrt{D}$ distribution for 2017 data, the mean value is $<$$\sqrt{D}$$>\simeq$0.25 ns.
  For the definition of $Pull$ and $D$ see text.}
\end{figure}

%%%%%%%%%%%%%%%%%%
The single hit time resolution is estimated from the same positron sample
as for the event time resolution measurement.
In order to exclude the dependence on the number of hits, the $Pull$ variable is introduced as  
$Pull = 0.5\cdot(T_1 -T_2)\sqrt{N_{hits}}$.
Since $\sigma(0.5\cdot(T_1 - T_2)) \simeq \sigma(T_{RICH})$
(under the same assumption as in~\ref{subsec:event}) and
$\sigma(T_{RICH}) \simeq \sigma(T_{hit}) / \sqrt{N_{hits}}$,
$\sigma(Pull) \simeq \sigma(T_{hit})$,
where $\sigma(T_{hit})$ is
the single hit time resolution.
The $Pull$ distribution is shown in
figure~\ref{fig:single_hit_time_resolution}-left.
The gaussian fit gives
 $\sigma(T_{hit})$ = 0.24 ns. 
 The origin of non-gaussian tails is the same as for figure~\ref{fig:event_time_resolution}.
As a cross-check, the single hit time resolution is calculated from the time variance
% $D = <t_{hit}^2> - <t_{hit}>^2$. 
% $D = N_{hits} / (N_{hits} - 1) \cdot ( <t_{hit}^2> - <t_{hit}>^2 )$. 
$D = < N_{hits} / (N_{hits} - 1) \cdot ( t_{hit} - <t_{hit}>)^2 >$. 
The distribution of $\sqrt{D}$ is shown in 
figure~\ref{fig:single_hit_time_resolution}-right. Its mean value 
%(0.24 ns) 
(0.25 ns)
is an alternative estimate
of $\sigma(T_{hit})$  and is 
%equal
similar 
to the value obtained from the $Pull$ width.

Using the measured value of the average number of hits $<$$N_{hits}$$>$ = 14.1
(this value is higher than the one from Section~\ref{subsec:stability} due to the additional cut on the ring 
acceptance to exclude the cone intersection with the beam pipe, see~\cite{alignment} for details), 
one can estimate the event time resolution under the assumption of constant value of $N_{hits}$: 
% $\sigma(T_{event}) = \sigma(T_{hit}) / \sqrt{<N_{hits}>} \simeq 0.064$ ns
$\sigma(T_{event}) = \sigma(T_{hit}) / \sqrt{<N_{hits}>} \simeq 0.06$ ns
which is close to the measured value $\sigma(T_{RICH}) \simeq 0.07$ ns.

%%%%%%%%%%%%%%%%%%
%%%%%%%%%%%%%%%%%%
\section{Conclusions}
\label{sec:conclusions}
%%%%%%%%%%%%%%%%%%
%%%%%%%%%%%%%%%%%%
The RICH detector was installed at the beginning of the NA62 pilot run in 2014 and commissioned in 2014-2015. The physics runs took place in 2016-2018.

In this paper the details of the NA62 RICH light detection and read-out systems are described. The light detection efficiency is estimated for single PMs and the average value is found to be 10\%. 
%(??? da discutere). 
The time resolution is measured using the positron sample: the event time resolution is 0.07 ns, 
much better than the requested 0.1 ns resolution needed to cope with the experiment particle rate,
while the single hit time resolution is 0.24 ns.

The RICH detector has been crucial in the NA62 physics results, contributing to several aspects of the analysis:
the optimal RICH time resolution has been essential
for the L0 trigger and to suppress accidental background, 
while the PID performance has been indispensable to reduce background from kaon decays.

%%%%%%%%%%%%%%%%%%
\acknowledgments
%%%%%%%%%%%%%%%%%%

The construction of the RICH detector would have been impossible without the enthusiastic work of 
% many 
technicians from
the University and INFN of Perugia and Firenze and the staff of the CERN laboratory.
%and  the strong collaboration 
%with 
% Vito Carassiti from 
%INFN Ferrara. 
%A special thanks to the NA62 collaboration for the full dedication to 
%the construction, commissioning and running of the experiment.
We are grateful to the whole NA62 Collaboration for its support during the construction, installation and commissioning of the RICH detector  and for its dedication in operating  the experiment in data taking conditions.
The software framework and algorithms developed by the NA62 collaboration for off-line data processing were used to obtain the results presented here.

%%%%%%%%%%%%%%%%%%

\end{document}